\documentclass[%
reprint,
 amsmath,amssymb,
 aps,pre
]{revtex4-2}

\usepackage{color}
\usepackage{graphicx}% Include figure files
\usepackage{dcolumn}% Align table columns on decimal point
\usepackage{bm}% bold math
\usepackage{amsmath}
\let\vec\mathbf
\usepackage{mathtools}
\usepackage{float}

\newcommand{\eq}{Eq. }

\begin{document}

\title{Collective gradient sensing with limited positional information}

\author{Emiliano Perez Ipiña}
\email{emperipi@jhu.edu}
 \affiliation{Department of Physics \& Astronomy, Johns Hopkins University, Baltimore, MD.}
 
\author{Brian A. Camley}%
\email{bcamley@jhu.edu}
\affiliation{Department of Physics \& Astronomy and Biophysics, Johns Hopkins University, Baltimore, MD.}

\date{\today}% It is always \today, today,

\begin{abstract}
Eukaryotic cells sense chemical gradients to decide where and when to move. Clusters of cells can sense gradients more accurately than individual cells by integrating measurements of the concentration made across the cluster. Is this gradient sensing accuracy impeded when cells have limited knowledge of their position within the cluster, i.e. limited positional information? We apply maximum likelihood estimation to study gradient sensing accuracy of a cluster of cells with finite positional information. If cells must estimate their location within the cluster, this lowers the accuracy of collective gradient sensing. We compare our results with a tug-of-war model where cells respond to the gradient by a mechanism of collective guidance without relying on their positional information. As the cell positional uncertainty increases, there is a trade-off where the tug-of-war model responds more accurately to the chemical gradient.  However, for sufficiently large cell clusters or shallow chemical gradients, the tug-of-war model will always be suboptimal to one that integrates information from all cells, even if positional uncertainty is high. 
\end{abstract}

%\keywords{}

\maketitle

\section{Introduction}
Eukaryotic cells may directionally migrate by sensing external cues of many types -- mechanical, electrical, topographical, or chemical \cite{sengupta2021principles}, and in these processes their accuracy is often limited by basic physical principles \cite{levine2013physics}.
The best-known of these processes is chemotaxis, where cells 
sense and follow gradients in concentrations of chemical cues -- an example of gradient sensing.  
Gradient sensing and directed migration are essential to many fundamental biological processes, such as immune response \cite{afonso2012ltb4}, embryonic development \cite{montell2012group,theveneau2010collective}, and cancer metastasis \cite{stuelten2018cell,roussos2011chemotaxis}.
In these processes, cells may migrate individually or collectively -- in sheets, streams, or small clusters \cite{mishra2019cell,friedl2009collective,haeger2015collective}. 
Cell cluster migration may be particularly important in cancer, where small clusters of tumor cells are associated with more harmful metastasis~\cite{cheung2016collective}.
There are several advantages for cells to migrates as groups~\cite{mayor2016front}, one of these being gradient sensing.
In several biological systems \cite{theveneau2010collective,malet2015collective,ellison2016cell}, groups of cells can chemotax where single cells fail to do so -- cells work collectively to improve their ability to sense a gradient \cite{camley2018collective}. 

How this enhanced sensing arises is not fully understood, and may not have a universal mechanism \cite{camley2018collective}, but two types of cell clusters seem to primarily sense using measurements of the concentration at the cluster edge. 
In clusters of neural crest cells responding to gradients of Sdf1, edge cells are polarized out from the cluster center by contact inhibition of locomotion \cite{theveneau2010collective}, suggesting a model of chemotaxis as a tug-of-war by the perimeter cells \cite{camley2016emergent}. 
A similar edge-driven mechanism is observed in gradient-sensing lymphocyte clusters~\cite{malet2015collective,copenhagen2018frustration}.
These observations suggest that in these cell types only edge cells are involved in cluster chemotaxis.
Why use only the edge cells to sense the gradient instead of all of them? Our initial intuition is that using only the edge cells wastes information \footnote{This assumes that interior cells can sense the concentration. In three-dimensional organoids, interior cells are isolated and may not have an equal ability to measure concentration.}.
In earlier work, we found a fundamental limit on gradient sensing accuracy \cite{camley2017cell}, observing that collective chemotaxis was likely limited by cell-to-cell variability. 
In that model, measurements from all cells and their relative positions are used to compute the best estimate of the gradient direction. %

Why would clusters use sensing mechanisms driven by edge cells, like the tug-of-war mechanism, instead of the optimal use of all cells? One hypothesis is that the benefit of the extra information is small. Another hypothesis is that there are relevant sources of noise that are not included in Ref. \cite{camley2017cell}. 
A major assumption implicit in our earlier work \cite{camley2017cell} is that cells ``know'' their position within a cluster -- the best estimator of the chemical gradient involves a sum weighted by a cell's position relative to the cluster center.
Cells can measure their position within an embryo or aggregate by measuring diffusible \cite{gregor2007probing} or mechanical \cite{nunley2021generation} signals but this positional information is limited ~\cite{wolpert1989positional,wolpert2011positional,tkavcik2021many,gregor2007probing}. 
Cells in {\it Drosophila} embryos can measure their position to the order of a cell length~\cite{gregor2007probing}, using temporal and possibly spatial averaging of graded signals~\cite{gregor2007probing, dubuis2013positional,ipina2016fluctuations}. 
In this paper, we explore the optimal strategies for collective sensing of gradients when cells have limited information on their position in the cluster, showing that positional uncertainty reduces gradient sensing accuracy. We also compare our model to an extension of the tug-of-war model of \cite{camley2016emergent}, where only edge cells respond to chemoattractant and no positional information is required. We find that these tug-of-war strategies are optimal for small clusters or strong gradients. However, for a sufficiently large cluster the benefits of using all cells' data will eventually outweigh the error from finite positional information.

\section{Models and Results}

\subsection{Maximum likelihood estimates of gradient with limited positional information}
We consider $N$ circular cells of radius $R_\mathrm{cell}$ arranged in 2D clusters under a linear chemoattractant gradient $\vec{g} = g\left(\cos(\phi), \sin(\phi)\right)$, see Fig.~\ref{fig:figure1}a. $g$ and $\phi$ are the gradient magnitude or steepness and direction respectively. 
The chemoattractant concentration at cell $i$ is $c(\vec{r}_i) = c_0(1 + \vec{g}(\vec{r}_i-\vec{r}_\textrm{cm}))$, where $\vec{r}_i$ is the cell's position, and $\vec{r}_\textrm{cm}$ is the position of the cluster center of mass, computed as $\vec{r}_\textrm{cm} = \frac{1}{N}\sum \vec{r}_i$.
$c_0$ is the chemoattractant concentration at the cluster center of mass.
Cells perform a collective measurement of the gradient by integrating individual cells' measurements of local chemoattractant concentration, which we denote by $M_i$.
Cells sense the concentration by binding the chemoattractant molecules to receptors at their surface, which is subject to fluctuations due to the molecules' diffusion in reaching receptors, and intrinsic ligand-receptor kinetics~\cite{berg1977physics,bialek2005physical,kaizu2014berg,ipina2016fluctuations}.
The reading concentration error of a cell at position $\vec{r}_i$ with $n_r$ receptors that bind ligand molecules with dissociation constant $K_D$, can be expressed as ${\delta c}^2(\vec{r}_i) = \frac{c(\vec{r}_i)}{n_r K_D}(c(\vec{r}_i) + K_D)^2$~\cite{camley2017cell, ipina2016fluctuations}.
In addition, even genetically identical cells do not respond in the same way to the chemoattractant due to cell-cell variability (CCV) resulting from fluctuations in their internal molecular machinery~\cite{camley2017cell}. 
Then, the concentration measurement of cell $i$ is,
\begin{align}
    M_i &= \frac{c(\vec{r}_i) + \delta c(\vec{r}_i)\eta_i}{\overline{c}} + \sigma_\Delta\xi_i,
    \label{eq:M_i}
\end{align}
where $\eta_i$ and $\xi_i$ are independent Gaussian noises with mean value 0 and variance 1,  $\overline{c}$ is the mean concentration over the cluster, and $\sigma_\Delta$ is the CCV standard deviation. 
Note that $\overline{c} = \frac{1}{N}\sum_i^N c(\vec{r}_i) = c_0$.
For typical values of $\sigma_\Delta \ge 0.05$ (i.e. 5\% cell-to-cell variability), ligand-receptor noise is less critical than CCV noise \cite{camley2017cell}.

The gradient sensing error can be obtained by combining the measurements in Eq.~(\ref{eq:M_i}) through the maximum likelihood estimation (MLE) method, as in~\cite{camley2017cell}. 
However, this implicitly assumes that the positions of the cells are known, as each measurement $M_i$ correspond to a given position $\vec{r}_i$ within the gradient. 
Here we extend~\cite{camley2017cell} by assuming that cells have limited positional information.
As a result, cells need to estimate their positions, $\vec{r}^*_i$, in addition to the local concentration.
We assume that cells get their positional information from an independent process different from the sensing concentration such that $\vec{r}^*_i$ and $M_i$ are independent.
Thus, the measured position for the cell $i$ is,
\begin{align}
    \vec{r}_i^* &= \vec{r}_i + \mathcal{A}_i\bm{\xi}_{\vec{r}_i},
    \label{eq:ri}
\end{align}
where $\mathcal{A}_i$ is a matrix such that $\vec{\Sigma}_i = \mathcal{A}_i\mathcal{A}_i^T$ is the covariance matrix of the positional errors, and $\bm{\xi}_{\vec{r}_i}$ is a two dimensional vector of uncorrelated normal distributed numbers. $X^T$ denotes the transpose of $X$.

Following the same procedure as in~\cite{camley2017cell}, we apply the maximum likelihood estimator (MLE) method, to compute the gradient sensing error limits. 
We start by writing the probability distribution for a cell $i$ to measure a signal $M_i$ and a position $\vec{r}^*_i$ given the gradient $\vec{g}$ and its true position $\vec{r}_i$, $p(M_i,\vec{r}_i^*|\vec{g},\vec{r}_i)$. 
As $M_i$ and $\vec{r}^*_i$ are independent, this probability function can be broken into the product of probabilities $p(M_i,\vec{r}_i^*|\vec{g},\vec{r}_i)=p(M_i|\vec{g},\vec{r}_i) p(\vec{r}_i^*|\vec{r}_i)$.
Given that Eq.~(\ref{eq:M_i}) takes the sum of two uncorrelated Gaussian random variables, which is also Gaussian, then, the probability distribution for $M_i$ is
\begin{equation}
    \label{eq:p_Mi}
    p(M_i|\vec{g},\vec{r}_i) = \frac{1}{\sqrt{2\pi h_i}}\exp{\left[-\frac{\left(M_i-\mu_i\right)^2}{2h_i}\right]},
\end{equation}
where $\mu_i = 1 + \vec{g}\cdot\delta\vec{r}_i$, and $h_i=(\delta c_i/c_0)^2 + \sigma^2_\Delta$.
Next, the probability distribution that a cell $i$ measures a position $\vec{r}^*_i$ given the true position $\vec{r}_i$ is,
\begin{equation}
    p(\vec{r}_i^*| \vec{r}_i) = \frac{1}{\sqrt{(2\pi)^2 \vert \bm{\Sigma}_i\vert}} \exp{\left[ -\frac{(\vec{r}_i^* - \vec{r}_i)\bm{\Sigma}_i^{-1}(\vec{r}_i^* - \vec{r}_i)^T}{2}\right]}.
\end{equation}
The likelihood of parameters $\vec{g},\{\vec{r}_i$\} given that the clusters measures values $\{M_i\}$ and $\{\vec{r}^*_i\}$ is $\mathcal{L}(\vec{g},\{\vec{r}_i\}|\{M_i, \vec{r}_i^*\}) \equiv p(\{M_i, \vec{r}_i^*\} | \vec{g},\{\vec{r}_i\})$. 
Assuming the measurement each cell performs is independent, this likelihood then factorizes into a product over cells $i$,  $\mathcal{L}(\vec{g},\vec{r}_i|\{M_i, \vec{r}_i^*\})=\prod_i p(M_i|\vec{g},\vec{r}_i) p(\vec{r}_i^*|\vec{r}_i)$.
Estimators of the gradient magnitude $g$ and orientation $\phi$ can be obtained by maximizing this likelihood. However, we are more interested in the best possible accuracy for an unbiased estimator $\hat{g}$, which is given by the Cram\'er-Rao bound \cite{kay1993fundamentals},
\begin{align}
    \sigma_g^2 \equiv \langle (g-\hat{g})^2\rangle = \left(\mathcal{I}^{-1}\right)_{g,g},
    \label{eq:g_def_error_fisher_inf}
\end{align}
where $\mathcal{I}^{-1}$ is the inverse of the Fisher information matrix, given by $\mathcal{I}_{\alpha,\beta} = -\left\langle \frac{\partial^2\ln\mathcal{L}}{\partial \alpha\partial \beta}\right\rangle$, with $\alpha$ and $\beta$ are parameters of the likelihood function, {\it i.e.} $\vec{g}$ and $\{\vec{r}_i\}$.

It is possible to analytically compute the Fisher information $\mathcal{I}$ and thus the best possible accuracy for measuring the chemical gradient in the limit of positional information in terms of a sum over cell positions and the positional information at each location. These results are presented in full detail in Appendix \ref{sec:MLE_appendix}. However, the results are much more intuitively understandable with a few key assumptions.

\subsection{Simplest case: shallow gradients and constant positional error}

\begin{figure}[!ht]
    \centering
    \includegraphics[width=0.48\textwidth]{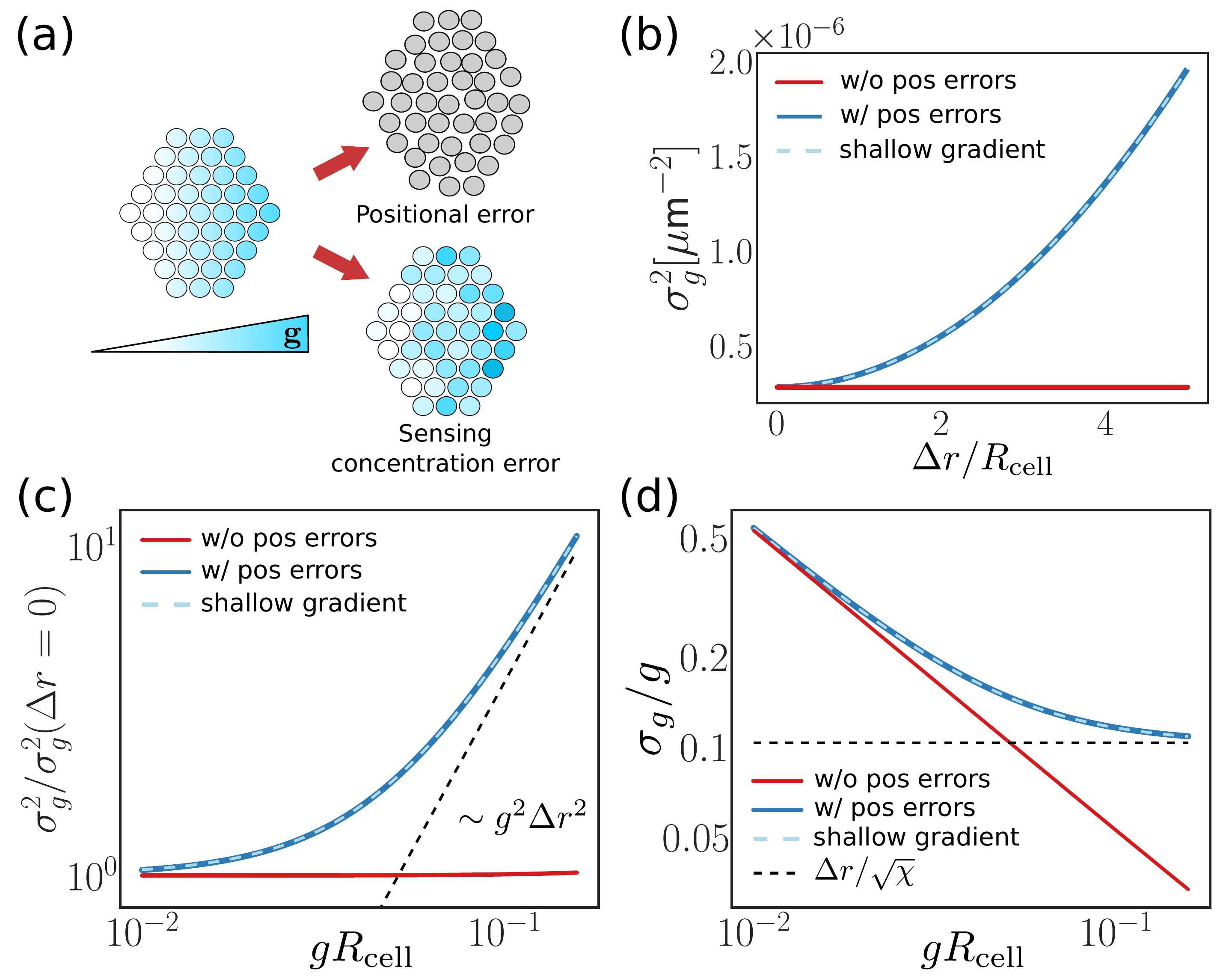}
    \caption{(a) Cluster under a linear gradient and a representation of the two sources of uncertainties: sensing concentration and positional errors. (b) Gradient sensing error increases with positional uncertainty. (c)--(d) \textbf{Positional uncertainties becomes more dominant in steeper gradients.} (c) Gradient sensing error normalized by its value in the absence of positional errors, $\sigma_g^2 (\Delta r=0)$, as a function of the gradient steepness (blue solid line). The black dashed-line indicates the limit where the gradient sensing error is dominated by positional errors and it follows $\propto g^2\Delta r^2$. (d) Gradient sensing relative error $\sigma_g / g$ as a function of the gradient steepness. Parameters: number of cells $N = 37$ (corresponding to an hexagonal cluster of $Q=3$ layers), $R_\mathrm{cell}=10\mu\mathrm{m}$, (b) $g=0.005\mu\mathrm{m}^{-1}$, (c)--(d)  $\Delta r= 10\mu\mathrm{m}$.}
    \label{fig:figure1}
\end{figure}

Cells will likely have different positional uncertainties, depending on the underlying process through which cells get their positional information, such as an external or a self-generated signal. 
To gain understanding, we will first look at the simplest case by assuming all the cells have the same positional covariance matrix, $\vec{\Sigma}_i=\vec{\Sigma}$, which we also assume is isotropic $\vec{\Sigma}_{ls} =\Delta r^2\delta_{ls}$, where $l,s$ are matrix element indexes. (We study clusters with varying positional uncertainty in Section \ref{sec:varying_pi}). Within this section, the limited positional information is only characterized by a single number, $\Delta r$ -- the error in measuring a cell's position.
In addition to assuming a constant isotropic error, we simplify our results by assuming that the gradient is relatively shallow -- i.e. that each cell has the same reading concentration errors, $\delta c_i^2 \approx \overline{\delta c}^2 = \frac{c_0\left(c_0+K_D\right)^2}{n_r K_D}$. 
Finally, we consider that our cluster of cells is roughly circular: the isotropy of the problem then makes $\sum_i \delta r_{x_i}^2\approx \sum_i\delta r_{y_i}^2\approx\frac{1}{2}\sum_i\vert\delta\vec{r}_i\vert^2$, and that $\sum_i \delta r_{x_i}\delta r_{y_i}\approx 0$. (In practice, for the calculations we present here, we show results for hexagonally packed clusters of layers of cells, as in \cite{camley2016emergent,camley2017cell}).
With all this, we can simplify the results of Appendix \ref{sec:MLE_appendix} to find the best possible error in measuring the gradient magnitude $g$ as: 
\begin{align}
    \sigma_g^2 \approx& \frac{1}{\chi} \left(\bar{h} + \left(g\Delta r\right)^2\right), \label{eq:sigma_g2_SGA},
\end{align}
where $\bar{h} = \overline{\delta c}^2/c_0^2 + \sigma_{\Delta}^2$ and $\chi = \frac{1}{2} \sum_i \vert \delta \vec{r}_i\vert^2$. 
Moreover, we find that the gradient direction error is equal to the gradient magnitude relative error,
\begin{equation}
\sigma_\phi = \frac{\sigma_g}{g}.\label{eq:sigma_phi2_SGA}
\end{equation}
See Appendix~\ref{sec:MLE_appendix} for detailed derivation of Eqs.~(\ref{eq:sigma_g2_SGA})--(\ref{eq:sigma_phi2_SGA}).

How does the presence of positional uncertainties affect gradient sensing?
In Fig.~\ref{fig:figure1}(b) we show that the gradient sensing error, $\sigma_g^2$, as a function of the positional uncertainty $\Delta r$, using the generic solution of Eq.~(\ref{eq:g_def_error_fisher_inf}) (see Appendix~\ref{sec:MLE_appendix}) and the shallow gradient approximation from Eq.~(\ref{eq:sigma_g2_SGA}). 
Unsurprisingly, positional uncertainties increase gradient sensing error -- the more uninformed cells are about their positions, the worse an estimate the cluster makes.
This added error from positional uncertainty increases as $\Delta r^2$ in the shallow gradient limit (Eq.~\ref{eq:sigma_g2_SGA}). We see in Fig.~\ref{fig:figure1} that the additional uncertainty from finite positional information is significant when $\Delta r$ is of order of the cell size. 

Eq.~(\ref{eq:sigma_g2_SGA}) also tells us there are two terms controlling the estimate error. 
The first, $\bar{h}$, is the contribution from the cells reading and reporting their local concentration. 
The second term, is the positional uncertainty contribution to the gradient sensing error, controlled by $g\Delta r$. 
The relative weight of these two terms allows to identify two regimes in which the gradient sensing is limited in one case by the reading concentration fluctuations, $(g\Delta r)^2\ll \bar{h}$, and on the other case by the positional uncertainties, $(g\Delta r)^2\gg \bar{h}$. 
When positional uncertainties are smaller than reading concentration errors, we recover the results from~\cite{camley2017cell}, $\sigma_g^2 \approx \frac{1}{\chi}\bar{h}$. In the other extreme, $(g\Delta r)^2\gg \bar{h}$, gradient sensing is primarily limited by positional information, $\sigma_g^2 \approx \frac{1}{\chi}g^2\Delta r^2$. 

The gradient steepness $g$, controls the relative importance of positional information: the steeper the gradient is, the larger the contribution from the positional uncertainty.
In Fig.~\ref{fig:figure1}(c) we show the gradient sensing error $\sigma_g^2$ normalized by the bound in absence of positional uncertainty $\sigma_g^2(\Delta r = 0)$ increases with the gradient steepness $g$. 
In steeper gradients, small errors in measuring a cell's position lead to larger differences between the concentration at the true position, $c(\vec{r}_i)$, and at the estimated position, $c(\vec{r}^*_i)$. 
On the other hand, this also implies, that in the limit of $g\rightarrow 0$, where error bounds are most important (where single cells fail to follow a gradient while clusters do), that positional information does not play a significant role. 

In analogy to single-cell gradient sensing, it would be natural to assume that the relative error, $\sigma_g/g$, which controls the angular uncertainty $\sigma_\phi$, always gets better with the increasing $g$ -- i.e. that cells can better measure the angle to the gradient direction if the gradient is steeper. This is true for clusters without positional error \cite{camley2017cell}: $\sigma_g/g$ is plotted as the red line in Fig.~\ref{fig:figure1}(d).
However, in the presence of positional uncertainties, the relative error is limited at large $g$, converging to a constant value $\Delta r/\sqrt{\chi}$ in the case of a shallow gradient (Fig.~\ref{fig:figure1}(d)).
This surprise occurs because the increasing error due to positional uncertainty $\sigma_g^2 \sim g^2$ in Eq.~(\ref{eq:sigma_g2_SGA}) exactly balances the $g^2$ in the denominator of $\sigma_g^2/g^2$.

\subsection{Non-uniform positional information}
\label{sec:varying_pi}
\begin{figure}[!ht]
    \centering
    \includegraphics[width=0.48\textwidth]{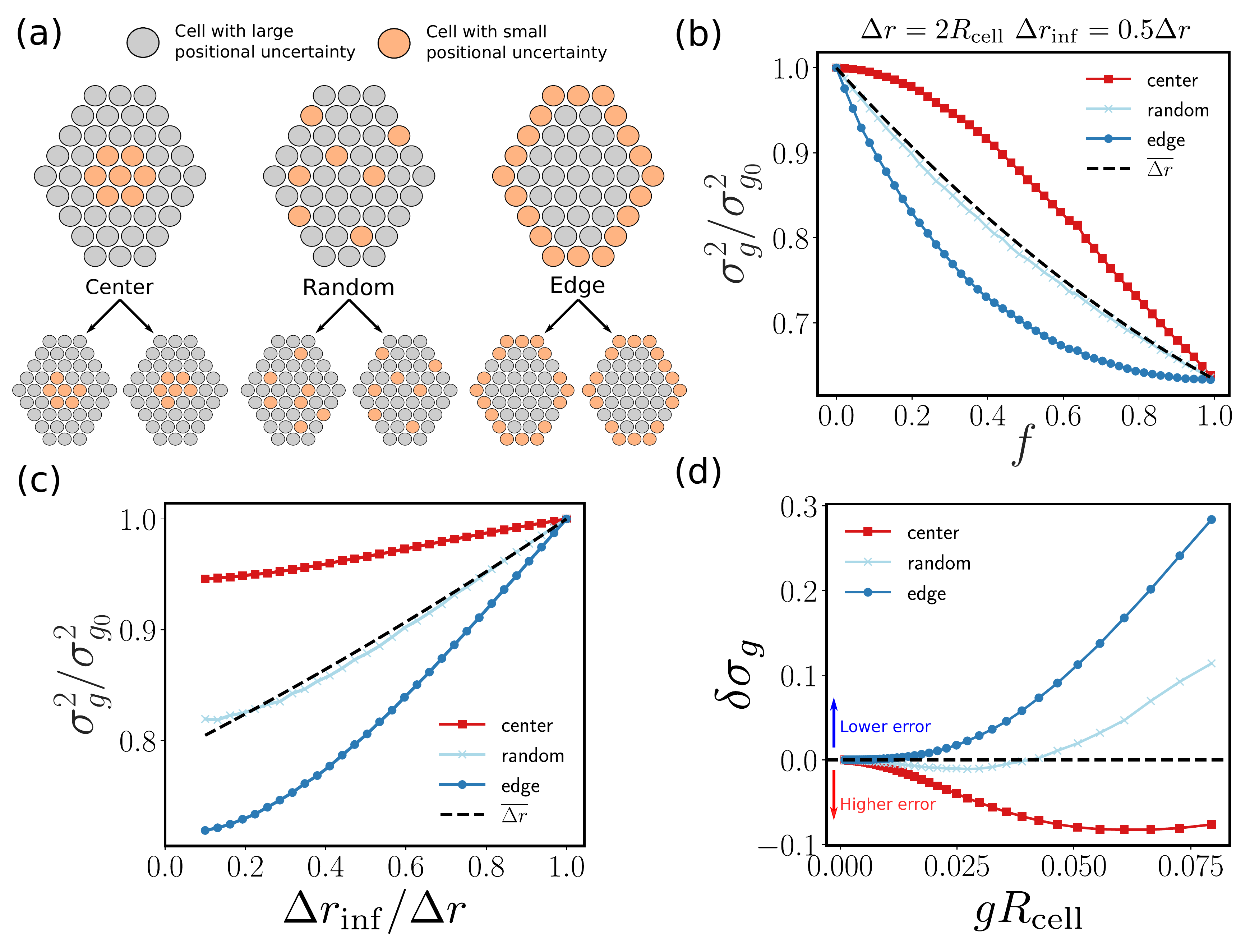}
    \caption{Gradient sensing errors for a cluster with a fraction $f$ of informed cells. (a) Informed cells are distributed over the cluster in three distinct manners: closer to the cluster's center, randomly, or closer to the cluster's edge. The row with the small clusters illustrates the existence of multiple equivalent ways of distributing the informed cells. (b)--(c) Gradient sensing variance, $\sigma_g^2$, normalized by the case of non informed cells, $\sigma^2_{g_0} = \sigma^2_g(f=0)$ as a function of the fraction of informed cells (b) and the ratio of the two types of cell positional uncertainties (c). (d) Relative difference between informed cluster and the equivalent constant-positional-error cluster, $\delta \sigma_g = \frac{\sigma_g(\overline{\Delta r})-\sigma_g(\Delta r_{\mathrm{inf}}, \Delta r)}{\sigma_g(\Delta r_{\mathrm{inf}}, \Delta r)}$, as a function of the gradient steepness. $\delta \sigma_g>0$ ($\delta \sigma_g<0$) indicates a lower (higher) gradient sensing error for the cluster with the informed cells compared to the uniform positional error cluster. The dashed black line indicates $\delta \sigma_g=0$. For (b)--(c)--(d) color codes are the same. Solid lines with symbols represent the three different distribution patterns: Center (red squares), Random (light blue crosses), and Edge (dark blue circles). The dashed black line represents the equivalent one type of cells cluster with positional uncertainty equal to $\overline{\Delta r} = f\Delta r_{\mathrm{inf}} + (1-f)\Delta r$. Parameters: cluster size $N=91$ ($Q=5$ layers in the hexagonal cluster), $\Delta r = 2R_\mathrm{cell}$, (b) $\Delta r_{\mathrm{inf}}= 0.5\Delta r$, (c) $f=0.25$, (d) $f=0.25$, $\Delta r_{\mathrm{inf}}= 0.2\Delta r$, and $\overline{\Delta r}=20\mu\mathrm{m}$. Curves show an average of 100 realizations.}
    \label{fig:figure2}
\end{figure}

In the above section we assumed that all cells have the same positional errors. 
However, positional uncertainty is not necessarily uniform within the cluster. If cells get their positional information from a secreted factor or mechanical signal, the distance from the source may influence the accuracy $\Delta r$ (see, e.g. \cite{alon2019introduction}).
In addition, obtaining positional information may have an associated cost and therefore, it may be optimal for the cell cluster to only have some cells measure their location. 
We study a prototypical example of both cell specialization and positional errors varying from cell to cell by introducing a second type of cells which have more positional information.
These ``informed'' cells have a positional error $\Delta r_{\mathrm{inf}}$ that is smaller than ``normal'' cells, $\Delta r_{\mathrm{inf}} \leq \Delta r$. 
Then, we distribute a fraction $f$ of informed cells at different positions over the cluster to see the effect of positional uncertainty localization. 
We follow three different types of distributions: informed cells are distributed (i) at the edge, (ii) at the center, or (iii) randomly over the cluster, see Fig.~\ref{fig:figure2}(a).
We use the general solution for the gradient sensing error given by Eqs.~(\ref{eqs:mle_general_solution_x})--(\ref{eqs:mle_general_solution_xy}).
We note that there are often many cells that are equidistant from the cluster center or edge, so unless a layer of the hexagonal packing is completely filled, there will be multiple possible configurations at a particular fraction of informed cells $f$ (Fig. \ref{fig:figure2}a, bottom row). To address this, we sample 100 realizations for each distribution. Fig~\ref{fig:figure2} shows an average over these realizations.
Gradient sensing has the lowest error when informed cells are closer to the edges and the largest when they are placed closer to the center of mass. 
This is illustrated in Fig.~\ref{fig:figure2}(b), that shows the gradient sensing error normalized by the case $f=0$, ($\sigma^2_{g_0} = \sigma^2_g(f=0)$), as a function of the fraction of informed cells. 
For all three types of distributions, increasing the fraction of informed cells reduces the error of $g$.
Gradient sensing error takes the same value for the three distributions, in the extreme cases of no informed cells ($f=0$) and all cells informed ($f=1$), but following different paths in between.
Informed cells distributed close to the edge lead to the lowest sensing errors of the three distributions.
Similarly, reducing the positional uncertainty of the informed cells results in higher accuracy in gradient sensing, which is again greater when informed cells are distributed closer to the edge (Fig.~\ref{fig:figure2}(c)).

To compare between clusters with informed cells and cluster with a single cell type population, we introduce the ``equivalent'' constant-positional-error cluster, in which all the cells have $\overline{\Delta r} = (1-f)\Delta r + f\Delta r_{\mathrm{inf}}$ (dashed black line in Fig.~\ref{fig:figure2}(b)--(d)).
Results show that placing informed cells closer to the edge enhances gradient sensing when compared to the equivalent cluster, while doing it closer to the cluster's center or randomly results in a worse or similar performance, respectively.
To further explore the role of informed cells and their distribution inside the cluster, in Fig.~\ref{fig:figure2}(d) we shows the relative difference between the gradient sensing errors for the cluster with informed cells and its constant-positional-error equivalent, $\delta \sigma_g= \frac{\sigma_g(\overline{\Delta r})-\sigma_g(\Delta r_{\mathrm{inf}}, \Delta r)}{\sigma_g(\Delta r_{\mathrm{inf}}, \Delta r)}$, as a function of the gradient steepness. 
Results show that increasing the gradient steepness enhances the trends described before, in which the benefit of reduced positional error is larger for cells at the cluster edge. 
This observation agrees with our previous results that showed that positional errors have a higher effect over gradient sensing in steeper gradients. 
Interestingly, in steeper gradients, even randomly distributed informed cells perform better than the equivalent cluster (light blue line in Fig.~\ref{fig:figure2}(d)). 
The reason is that by randomly distributing the informed cells, a few will lie close to the edge of the cluster, and those are the ones that dominate and end up having a major weight in sensing the gradient.

Together, these results show that positional information becomes more relevant at the edge of the cluster and is even more significant in steeper gradients. The importance of the edge is large enough that even having a few informed cells randomly localized to the edge can significantly decrease error.
If determining cell position is difficult, our results suggest that instead of maximizing positional information evenly across the entire cluster, cells may improve in gradient detection by prioritizing the localization of edge cells.
\\

\begin{figure}
    \centering
    \includegraphics[width=0.48\textwidth]{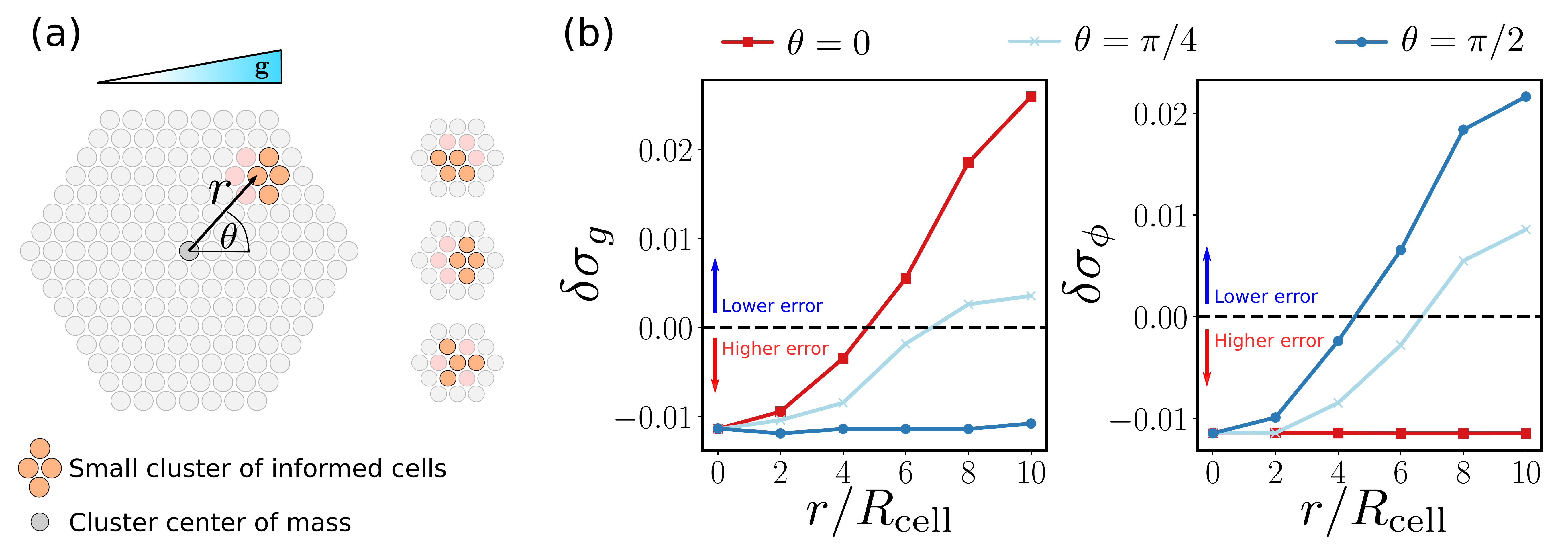}
    \caption{(a) Schematic cell cluster with a group of four informed cells located at a distance $r$ from the cluster's center and direction $\theta$ with respect to the chemoattractant gradient. On the Right: A non exhaustive representation of different configurations for the small cluster. (b) Relative gradient sensing steepness (left panel) and direction, $\delta \sigma_\phi = \frac{\sigma_\phi(\overline{\Delta r})-\sigma_\phi(\Delta r_{\mathrm{inf}}, \Delta r)}{\sigma_\phi(\Delta r_{\mathrm{inf}}, \Delta r)}$,  (right panel) errors when the informed group of cells is displaced in different directions: $\theta=0$ (red line and squares), $\theta=\pi/4$ (light blue line and crosses) and $\theta=\pi/2$ (dark blue line and circles). Same as in Fig.~\ref{fig:figure2}, $\delta \sigma_{g,\phi}>0$ ($\delta \sigma_{g,\phi}<0$) represent a lower (higher) error relative to the constant-positional-error equivalent cluster. The dashed black line indicates $\delta \sigma_{g,\phi}=0$ and also represent the equivalent cluster without informed cells and positional uncertainty $\overline{\Delta r}$. Parameters: cluster size $N=91$ and $\Delta r_{\mathrm{inf}}= 0.5\Delta r$. Curves show an average of 100 realizations.}
    \label{fig:figure3}
\end{figure}

We argued that informed cells are generally most beneficially placed at the cluster edge. How does gradient sensing accuracy change if we place these cells anisotropically? 
We study the change in the errors for gradient magnitude $g$ and direction $\phi$ when a small group of four ``informed cells'' are displaced from the center to the edge of the cluster in different directions, see Fig.~\ref{fig:figure3}(a). We note that for this case, $\sigma_\phi \neq \sigma_g / g$, due to anisotropy; we use the full analytical solution of Appendix \ref{sec:MLE_appendix} to compute the bounds $\sigma_g$ and $\sigma_\phi$. 
As in the previous case, there are multiple equivalent ways to place a cluster of informed cells at a position $(r,\theta)$, see Fig.~\ref{fig:figure3}(a). To account for such variability, we averaged over 100 realization for each condition pair $(r,\theta)$.
We find that the errors for the estimators of the gradient magnitude and direction evolve differently depending on the orientation of this cluster of cells, Fig.~\ref{fig:figure3}(b).
Moving the small informed cluster along the axis parallel to the gradient direction, $\theta=0$,  leads to an improvement in sensing the gradient steepness $g$.
By contrast, moving along the axis perpendicular to the gradient, $\theta=\pi/2$, sensing the orientation of the gradient is improved.   
Finally, there is a compromise situation when moving along $\theta = \pi/4$, where both gradient steepness and direction sensing are improved, but to a lesser extent. 
These results support again that positional information is more significant at the edge of the cluster, but also that the orientation with respect to the gradient matters.
In chemotaxis, for instance, cells need to sense the direction rather than the steepness of the gradient.
Therefore, a cluster might strive to locate cells at the edge of the axis perpendicular to the gradient direction to enhance its ability to chemotax.
However, this could only work in the case that the cluster has previous knowledge of the gradient direction.
Our findings are similar in concept to the ideas reported in~\cite{hu2011geometry} for single cells, in which elongated cells sense the gradient steepness and direction with different accuracies depending on the orientation of the cell.
Note that in comparing Fig.~\ref{fig:figure3}(b) panels that $\delta\sigma_g(0)\neq \delta\sigma_\phi(\pi/2)$ and $\delta\sigma_g(\pi/4)\neq \delta\sigma_\phi(\pi/4)$. This is due to hexagonal clusters not having perfect circular symmetry, being more elongated, in our case, in the direction parallel to the gradient.

\begin{figure*}
    \centering
    \includegraphics[width=0.9\textwidth]{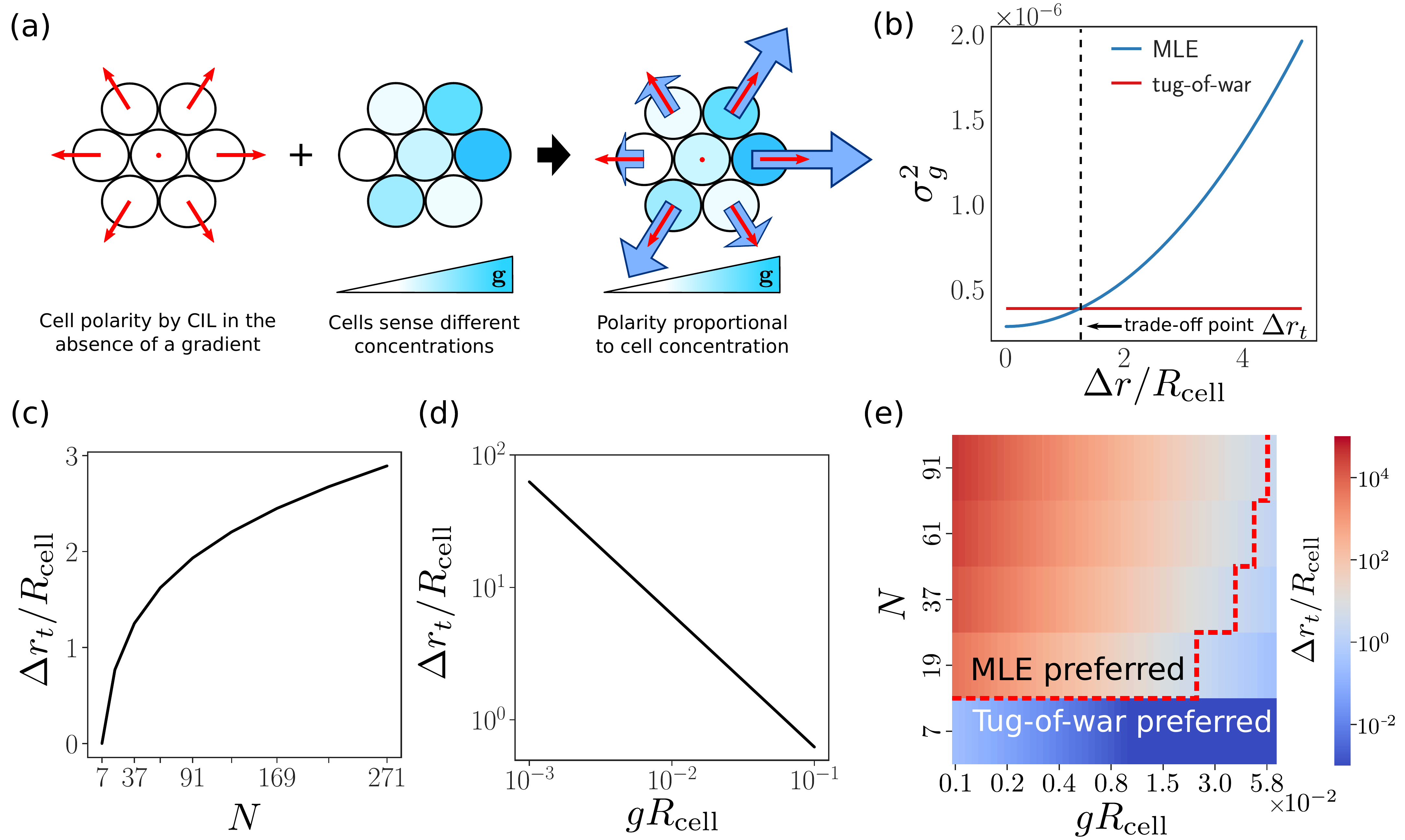}
    \caption{\textbf{Tug--of--war model} (a) Scheme representation of the tug--of--war model. (b) Gradient sensing variance as a function of the positional uncertainty for the tug--of--war model (red solid line) and the MLE model (blue solid line). The black dashed line marks the crossover and the trade-off positional uncertainty between the two models. (c)--(d) Trade--off value as a function of: (c) the cluster number of cells, and (d) the gradient steepness. (e) Phase diagram of the trade--off values for different cluster's sizes and gradient steepness. The dashed red line indicates a boundary given by $\Delta r_t = 2R_\mathrm{cell}$. Positional errors, $\Delta r$, and trade--off values, $\Delta r_t$, are re-scaled in of cell's radius. Parameters: (b) same as Fig.~\ref{fig:figure1}(b), (c) $g=0.005\mu\mathrm{m}^{-1}$, and (d) $N=37$ ($Q=3$).}
    \label{fig:figure4}
\end{figure*}
\subsection{Collective chemotaxis without positional information: tug--of--war model}
In our above approach, a group of cells senses a chemical gradient collectively by making local measurements of the chemoattractant concentration and and their positions.
Then, by applying the maximum likelihood estimator method, we find the best possible measurement of the gradient the group of cells can make. 
However, we showed that this measurement is constrained by the limited positional information available to the cells.
Here, we introduce a different mechanism by which cells sense and follow the gradient without needing cells to measure their locations, based on the model in~\cite{camley2016emergent}.
An interesting feature of this model, which makes it different from the previous one, is that it does not depend on the positions of the cells, but only on the direction to the nearest cells.
Cells interact by contact inhibition locomotion (CIL) \cite{camley2014polarity,stramer2017mechanisms}, in which contacting cells polarize away from each other. 
This interaction dynamic leads to a cluster where only cells at the edge are polarized \cite{theveneau2010collective,camley2016emergent,malet2015collective}, see Fig.~\ref{fig:figure4}(a). 
Cells on the inside of the cluster, completely surrounded by neighboring cells, do not polarize, whereas cells on the edges, which only see neighboring cells on the interior side, polarize away from the cluster.
Another important property of the model is that cell CIL is modulated by the local chemoattractant concentration. 
In this way, those cells at the front (up the gradient) of the cluster polarize and pull stronger than the ones on the back (down the gradient), resulting in a net movement toward up the gradient in a tug--of--war like dynamic.
From now on we will refer to this model as the ``tug--of--war'' model. 

Our goal in studying the tug-of-war model will be to extend the results of \cite{camley2016emergent} so that they can be directly compared with our maximum likelihood results above, which requires us to include fluctuations in concentration due to receptor-ligand binding as well as cell-to-cell variability. Following~\cite{camley2016emergent}, we start from the assumption that cells behave as stochastic particles.
The motility of cell $i$ results from the balance between the intracellular forces, $\vec{F}_{ij}$, due to cell-cell adhesion and volume exclusion, and the cell's polarity $\vec{p}_i$,
\begin{equation}
    \label{eq:cell_motility_tow}
    \partial_t \vec{r}_i = \vec{p}_i + \sum_{i\neq j}\vec{F}_{ij}.
\end{equation}
Computing the cluster velocity by summing over all cells in Eq.~(\ref{eq:cell_motility_tow}) and noting that $\vec{F}_{ij} = -\vec{F}_{ji}$, we find 
\begin{equation}
\label{eq:cluster_velocity}
    \vec{v}_c = \frac{1}{N} \sum_i \vec{p}_i.
\end{equation}

Cell polarity obeys the differential equation, 
\begin{equation}
\label{eq:pi_orstein-uhlembeck}
    \partial_t \vec{p}_i = -\frac{1}{\tau_p}\vec{p}_i + \sigma_p \bm{\epsilon}_i(t) + \beta_i\vec{q}_i,
\end{equation}
where $\tau_p$ is the characteristic time it takes for the polarity to relax to its steady state value, and $\sigma_p$ is the magnitude of a fluctuating noise that can drive the polarity away from its steady-state value.
$\bm{\epsilon}_i(t)$ is a vector Gaussian Langevin noise for cell $i$ that fulfills $\langle \bm{\epsilon}_i(t)\rangle=0$ and $\langle \epsilon_{i\mu}(t) \epsilon_{j\nu}(t')\rangle = \delta_{\mu\nu}\delta_{ij}\delta(t-t')$, with $i,j$ cell indexes, while $\mu,\nu$ are indexes for the Cartesian coordinates $x$, $y$. 

The last term in Eq.~(\ref{eq:pi_orstein-uhlembeck}) comes from the CIL interactions, where $\beta_i$ is the cell's susceptibility to CIL, and the vector $\vec{q}_i$ is the resulting direction of contact interaction of cell $i$ and its neighbors, $\vec{q}_i=\sum_{i\sim j} (\vec{r}_i-\vec{r}_j)/\vert\vec{r}_i-\vec{r}_j\vert$, where $i\sim j$ represent the sum over the cell's neighbors, defined as those cells within a distance 2.1$R_\mathrm{cell}$.
Note that $\vec{q}_i\approx 0$ for interior cells and $\vec{q}_i \neq 0$ for edge cells. 
The tug-of-war model of \cite{camley2016emergent} assumed that susceptibility of the cell $i$ is proportional to chemoattractant concentration, $\beta_i = \bar{\beta}\,c(\vec{r}_i,t)$. Here, we must handle the variability in the {\it measured} concentration due to both ligand-receptor binding and cell-to-cell variability. We assume that cells polarize in response to what they believe the chemoattractant concentration to be, $M_i\overline{c}$, instead of the true concentration value at the cell positions, $c(\vec{r}_i,t)$.
This will add an additional source of noise to Eq.~\ref{eq:pi_orstein-uhlembeck}, which we will now write explicitly. If the susceptibility of cell $i$ is $\beta_i(t) = \bar{\beta}c_0M_i(t)$, we can write $M_i(t) = c(\vec{r}_i,t)/c_0 + \Xi_i(t)$. Here, $\Xi_i(t) = \delta c_i/c_0 \zeta^c_i(t) + \sigma_\Delta \zeta^\Delta_i(t)$ is the noise in $M_i$. Note that, unlike Eq.~\ref{eq:M_i}, we must specify the time-dependence of the errors due to ligand-receptor binding and cell-to-cell variation. We explicitly introduce $\zeta^c_i(t)$ to measure the fluctuations due to concentration sensing and $\zeta^\Delta_i(t)$ for the fluctuations in cell-to-cell variability. 
We can characterize the time scale over which the errors due to receptor-ligand concentration and cell-to-cell variability are persistent by the correlation functions $\langle \zeta^c(t)\zeta^c(0) \rangle = C_c(t)$ and $\langle \zeta^\Delta(t)\zeta^\Delta(0) \rangle = C_\Delta(t)$.
Then we have $\langle \Xi_i(t)\rangle = 0$ and $\langle \Xi_i(t)\Xi_j(0)\rangle = \left(\delta c_i^2/c_0^2 C_c(t)+\sigma_\Delta^2 C_\Delta(t)\right)\delta_{ij}$.  
Next, we insert the expression for the polarity susceptibility $\beta_i(t) = \bar{\beta} c(\vec{r}_i) + \bar{\beta} c_0 \Xi_i(t)$,  into Eq.~(\ref{eq:pi_orstein-uhlembeck}) to arrive at
\begin{equation}
\label{eq:pi_orstein-uhlembeck_with_Mi_explicit}
    \partial_t \vec{p}_i =-\frac{1}{\tau_p}\left(\mathbf{p}_i - \gamma c(\vec{r}_i,t)\vec{q}_i\right) + \sigma_p\bm{\epsilon}_i(t) + \overline{\beta}c_0\vec{q}_i\Xi_i(t),
\end{equation}
where we have defined $\gamma = \tau_p\bar{\beta}$.

We can solve Eq.~\ref{eq:pi_orstein-uhlembeck_with_Mi_explicit} by directly integrating it. If we assume that $\tau_p$ is small enough for the polarity to relax to its steady-state solution before the cluster reorganizes, so the contact vectors $\vec{q}_i$ are fixed, then $\vec{p}_i$ is given by
\begin{align}
\label{eq:pi_sol}
\vec{p}_i(t) =& \bm{\mu}_i + \sigma_p\int_0^tdt' e^{-(t-t')/\tau_p}\bm{\epsilon}_i(t') \nonumber\\
 & + \bar{\beta}c_0\vec{q}_i\int_0^t dt' e^{-(t-t')/\tau_p}\Xi_i(t')
\end{align}
where $\bm{\mu}_i=\gamma c(\vec{r}_i)\vec{q}_i$ and we chose initial conditions $\vec{p}_i(0) = \bm{\mu}_i$.
From Eq.~(\ref{eq:pi_sol}), it is possible to compute the mean and variance of $\vec{p}_i(t)$. 
We assume that correlations decay exponentially, so $C_{s}(t) = e^{-t/\tau_s}$, with $\tau_s$ the correlation time and the index $s=\{c,\Delta\}$. Our final results will not be crucially dependent on $\tau_c$, $\tau_\Delta$, so this is primarily a convenience to make the analytical calculations simple.
Then, recalling that $\bm{\epsilon}_i(t)$ is a Gaussian uncorrelated noise with mean zero, we compute the mean and the covariance of $\vec{p}_i(t)$ (see Appendix~\ref{sec:app_tug_of_war}),
\begin{align*}
    \langle \vec{p}_i(t)\rangle =& \bm{\mu}_i,\\
\left\langle (p_{\mu_i}(t)- \langle p_{\mu_i}\rangle)(p_{\nu_i}(0)- \langle p_{\nu_i}\rangle)\right\rangle =& \frac{\sigma_p^2\tau_p}{2}\delta_{\mu\nu}\left(1-e^{-2t/\tau_p}\right)\\ 
+ \bar{\beta}^2q_{\mu_i}q_{\nu_i}\big(\delta c_i^2\mathrm{T}(t,\tau_c) &+  c_0^2\sigma_\Delta^2\mathrm{T}(t,\tau_\Delta)\big),
\end{align*}
where
\begin{align*}
    \mathrm{T}(t,\tau_s) =& \frac{\tau_p^2}{1+\upsilon_s}\Big(1 + \frac{1}{1-\upsilon_s}\\
    &\times\big((1+\upsilon_s)e^{-2t/\tau_p} - 2e^{-(1+\upsilon_s)t/\tau_p}\big)\Big),
\end{align*}
and $\upsilon_s = \frac{\tau_p}{\tau_s}$.
Note that typical values for the polarization relaxation time $\tau_p$ are  $\sim 20$ minutes \cite{camley2016emergent}. 
This is much longer than the typical correlation times for the chemoattractant--cell receptors binding dynamics which is on the order of the $\tau_c\sim 1$s \cite{camley2017cell}. 
However, recall that cell concentration sensing is limited by CCV which has a much slower correlation time than can reach up to 48 h in human cells~\cite{sigal2006variability,camley2017cell}.
To calculate the variance of $\vec{p}_i$ at a reasonable steady-state, then, we should think of the CCV noise as constant in time, and wait for times $t$ much longer than the other relaxation times. Explicitly, this means $\tau_\Delta \gg t \gg \tau_p \gg \tau_c$. In this limit, $\mathrm{T}(t,\tau_\Delta) \approx \frac{\tau_p^2}{1+\tau_p/\tau_\Delta}\approx \tau_p^2$ and $\mathrm{T}(t,\tau_c) \approx \frac{\tau_p^2}{1+\tau_p/\tau_c}$.
As a result, the polarity covariance is $\langle (p_{\mu_i}- \langle p_{\mu_i}\rangle)(p_{\nu_i}- \langle p_{\nu_i}\rangle)\rangle = V^i_{\mu\nu}$, where $V^i_{\mu\nu} = \left(\gamma^2c_0^2 h_{T_i} q_{\mu_i}q_{\nu_i} +\frac{\sigma_p^2\tau_p}{2}\delta_{\mu\nu}\right)$ and $h_{T_i} = \frac{\delta c_i^2/c_0^2}{1+\tau_p/\tau_c} +\frac{\sigma_\Delta^2}{1+\tau_p/\tau_\Delta} \approx \sigma_\Delta^2$. 
Essentially, the variance of the polarity has two terms: the time-averaged noise from measuring concentration, which is proportional to $h_{T_i}$ -- which we have named to make clear it is effectively a time average of $h_i$. The second arises from the added noise $\sigma_p$ in \eq \ref{eq:pi_orstein-uhlembeck}, which is a source of noise not included in the maximum likelihood model.

Then, with the distribution of polarities in hand, from Eq.~(\ref{eq:cluster_velocity}) we proceed to compute the expected value and variance of the cluster velocity. 
In the case of the expected value, pointing out that $\sum_i \vec{q}_i = 0$, we arrive at the following simple expression,
\begin{equation}
\label{eq:v_cluster_mean}
    \langle\vec{v}_c\rangle = \gamma c_0\mathcal{M}\vec{g},
\end{equation}
where $\mathcal{M}_{\mu\nu} = \frac{1}{N}\sum_i \delta r_{\mu_i}q_{\nu_i}$.
For the variance we have, 
\begin{equation}
\Delta v^2_{c_{\mu\nu}} = \frac{\gamma^2c_0^2}{N^2} \sum_i h_{Ti}\, q_{\mu_i}q_{\nu_i} + \frac{\tau_p\sigma_p^2}{2N}\delta_{\mu\nu}.   \label{eq:tug-of-war_v_cluster_errors}
\end{equation}

Now, we ask the question, what is the gradient sensing error for this tug--of--war model?
We use Eq.~(\ref{eq:v_cluster_mean}) and Eq.~(\ref{eq:tug-of-war_v_cluster_errors}) to propagate errors and compute the gradient sensing error for the tug--of--war model (Appendix \ref{sec:app_tug_of_war}),
\begin{equation}
\label{eq:gradient_sensing_errors_tug_of_war}
    \sigma_{g}^2 \approx \frac{2}{\left(\sum_i \vec{q}_i\cdot\delta \vec{r}_i\right)^2}  \left(\sum_i \vert \vec{q}_i\vert^2 h_{Ti}+ \frac{\sigma_p^2\tau_p}{\gamma^2c_0^2}N \right),
\end{equation}
under the assumption of isotropic clusters, as in Eq. \ref{eq:sigma_g2_SGA}. 

How can we make a consistent comparison between the tug-of-war model and the maximum likelihood estimation? There are two key differences. 
First, the tug-of-war model includes an additional source of noise -- the polarity noise $\sigma_p$ representing fluctuations in cell polarity arising from factors outside of concentration sensing. Second, the tug-of-war model, as a dynamical model, explicitly includes an average over a characteristic time $\tau_p$, while the maximum likelihood estimate is based on a single snapshot of the measured concentration.  
The first issue is simple to deal with: since the MLE method returns a lower bound for the gradient sensing error, it is most comparable to deal with the best possible situation for the tug--of--war model, {\it i.e.}, no polarity fluctuations, $\sigma_p=0$. 
The issue of time-averaging is less clear. Eq.~(\ref{eq:gradient_sensing_errors_tug_of_war})  depends on $h_{T_i} = \frac{\delta c_i^2/c_0^2}{1+\tau_p/\tau_c} +\frac{\sigma_\Delta^2}{1+\tau_p/\tau_\Delta}$. 
To compare with the instantaneous snapshot, we would have to choose $\tau_p \ll \tau_c, \tau_\Delta$, in which case $h_{T_i} \approx h_i$. However, this is inconsistent with our estimates above where $\tau_p \sim$ 20 minutes and $\tau_c \sim$ 1 second. 
In practice, however, this distinction is not very important, because both $h_{T_i}$ and $h_i$ are dominated by the CCV noise, so $h_{T_i} \approx h_i \approx \bar{h} \approx \sigma_\Delta^2$. We will then directly assume that $h_{T_i} = h_i = \bar{h}$, so that the tug-of-war and MLE results are directly comparable \footnote{An alternate approach would be to consider the maximum likelihood estimation of measured signals $M_i$ that have been time-averaged over a time $\tau_p$, in which case $h_i$ would be replaced in the MLE results by $h_{T_i}$. See Appendix \ref{sec:mle_time_average} for details.}. With these results, we can compute the uncertainty in the estimation of $g$ arising from the tug-of-war model as: 
\begin{equation}
    \sigma^2_g \approx\frac{1}{\chi_{\mathrm{tow}}}\,\, \bar{h},
    \label{eq:tug_of_war_error}
\end{equation}
where $\chi_{\mathrm{tow}} = \frac{\left(\sum_i \delta \vec{r}_i\cdot\vec{q}_i\right)^2}{2\sum_i \vert \vec{q}_{i}\vert^2}$. 
Note how similar Eq.~(\ref{eq:tug_of_war_error}) is to Eq.~(\ref{eq:sigma_g2_SGA}) when $\Delta r=0$. 
The only difference is in the geometrical factors $\chi$ and $\chi_{\mathrm{tow}}$. 
The prefactor $\chi_{\mathrm{tow}}$ involves sums over the CIL directional vectors $\vec{q}_i$, which are only non zero on the edges cells, supporting again that only those cells contribute to sensing the gradient in the tug--of--war model. 

We now want to compare the tug--of--war model and the MLE method to see which of them, and under what conditions, senses the gradient more accurately. 
Fig.~\ref{fig:figure4}(b) shows the gradient sensing error as a function of positional uncertainties for both models. 
In this case, since the tug--of--war model does not depend on positional uncertainties its gradient sensing error is constant. 
In absence of positional errors, the MLE method makes a better estimation of the gradient than the tug--of--war model, since this last one only uses information from the edge cells in contrast with the MLE model that uses all the cells. 
However, as cells become less informed about their positions, a trade-off occurs where, beyond a certain positional uncertainty value, $\Delta r_t$, the tug-of-war model turns out to be a better estimator of the gradient.
This crossover exists since the tug-of-war model does not need to estimate cell positions and is therefore independent of positional uncertainties. 
Under the shallow gradient approximation, from Eq.~(\ref{eq:sigma_g2_SGA}) and Eq.~(\ref{eq:tug_of_war_error}), we can find analytically the crossover between both models, 
\begin{equation}
\label{eq:Delta_r_trade_off_SGA}
    \Delta r_t \approx \frac{\sqrt{\bar{h}\left(\frac{\chi}{\chi_{\mathrm{tow}}}-1\right)}}{g}.
\end{equation}

Using maximum likelihood estimation, increasing the cluster's size can compensate for limitations in cells positional information, see Fig.~\ref{fig:figure4}(c).
This result can be easily predicted from Eq.~(\ref{eq:Delta_r_trade_off_SGA}), by considering the scaling of the geometrical factors $\chi$ and $\chi_{\mathrm{tow}}$ with the cluster radius, $R_{\mathrm{cluster}}$. 
We expect $\chi = \frac{1}{2} \sum_i \vert \delta \vec{r}_i\vert^2$ to scale as $\chi \sim R_\textrm{cluster}^4$ \cite{camley2017cell}. 
We find this by approximating the sum over cells as an integral over a roughly circular cluster, $\sum_i R_\textrm{cell}^2 \approx \int_{R_\textrm{cluster}} d^2 r$, so  $\chi =  \frac{1}{2} \sum_i \vert \delta \vec{r}_i\vert^2 \sim R_\textrm{cell}^{-2} \int_{R_\textrm{cluster}} d^2 r r^2 \sim R_\textrm{cluster}^4$. 
For the tug-of-war model, $\chi_{\mathrm{tow}} = \frac{\left(\sum_i \delta \vec{r}_i\cdot\vec{q}_i\right)^2}{2\sum_i \vert \vec{q}_{i}\vert^2}$. For a roughly circular cluster, $\vec{q}_i$ will be along the direction $\delta\vec{r}_i$ for cells at the edge, and zero for cells in the interior. 
Edge cells will have roughly the same magnitude of $\vec{q}_i$, $|q|$.
Thus, $\delta \vec{r}_i\cdot\vec{q}_i \approx |q| R_\textrm{cluster}$ for edge cells, and $\chi_{\mathrm{tow}} \approx (N_\textrm{edge} |q| R_\textrm{cluster})^2 / 2 N_\textrm{edge} |q|^2 \sim N_\textrm{edge} R_\textrm{cluster}^2$. 
As, for a roughly circular cluster, we expect the number of cells on the edge $N_\textrm{edge}$ to be proportional to $2\pi R_\textrm{cluster}$, we see $\chi_\mathrm{tow} \sim R_\textrm{cluster}^3$. 
Then, $\chi/\chi_{\mathrm{tow}}\propto R_{\mathrm{cluster}}$ and we get from Eq.~(\ref{eq:Delta_r_trade_off_SGA}) that the 
$\Delta r_t$ shifts to larger values with the cluster size, ($\Delta r_t \propto R_{\mathrm{cluster}}^{1/2}$).

On the other hand, the gradient steepness plays an opposite role in shifting the trade-off values. 
In a steeper gradient, the MLE method becomes more sensitive to uncertainties in cell positions. 
In this scenario, the tug-of-war model benefits from this limitation in the MLE method and senses the gradient more accurately, even in the presence of small positional errors.
As a consequence, the trade-off values shift towards smaller positional uncertainties with the gradient steepness, see Fig.~\ref{fig:figure4}(d).
We can summarize everything we described above with the phase diagram for the trade--off values shown in Fig.~\ref{fig:figure4}(e). 
From the figure, we see that areas in blue represent conditions in which the trade--off values are small. In these areas, a group of cells would be more prompt to uses the tug--of--war method rather than the MLE to sense the gradient. 
On the other hand, red areas show large trade--off values where it might be optimal for cells to pay the cost of estimating their positions and follow the MLE method for gradient sensing. 
Lastly, it is interesting to note the case of $N=7$. This cluster size corresponds to the 1--layer hexagonal cluster, where almost all cells are edge cells, and there is only one interior cell. 
The interior cell is located at the cluster's center, so that $\delta \vec{r}=0$, and therefore has no contribution to the gradient sensing in the MLE method (recall that $\chi=\sum_i \vert\delta \vec{r}_i\vert^2/2$ in the shallow gradient, see also for the general solution Eqs.~(\ref{eqs:mle_general_solution_x})--(\ref{eqs:mle_general_solution_xy})).
Consequently, both models use the same cells to sense the gradient, except that the MLE approach is affected by positional uncertainty, whereas the tug-of-war is not.
For this reason, even small uncertainties in cell positions are sufficient for the tug--of--war model to outperform the MLE and become the best gradient estimator. 
In the case of the shallow gradient approximation, we have that $\chi=\chi_{\mathrm{tow}}$ and from Eq.~(\ref{eq:Delta_r_trade_off_SGA}) we obtain $\Delta r_t \approx 0$.
\\

\section{Discussion}
We have studied the role of limited positional information in collective gradient sensing. 
Within the context of gradient sensing, how cells get their positional information -- or if they estimate the gradient without this information -- is still an open question.
Positional information within a group of cells has been studied in the context of developmental biology, where an external morphogen signal determine cell's fate~\cite{wolpert1989positional,gregor2007probing,tkavcik2021many}; cells may also sense their position within a colony by sensing mechanical stresses \cite{nunley2021generation}.
Moving from a sensed chemical or mechanical signal to positional knowledge within a cluster is a complex multistage process~\cite{tkavcik2021many}. Consequently, attaining precise positional information is a time costly process, and there are essential tradeoffs between the rate at which a pattern can be established and its precision\cite{tran2018precision}.
Establishing a reproducible gradient is likely even more challenging for migratory cells in clusters \cite{theveneau2010collective,malet2015collective}, as cell position and number of cells in the cluster all evolve over time.
However, there is a potential candidate for a morphogen-like source of positional information within neural crest cell cluster chemotaxis: the complement fragment C3a, which acts as a ``co-attractant" \cite{carmona2011complement}. Earlier simulations show that the co-attractant C3a can be established in a graded fashion \cite{camley2016collective}. 
Here, we have suppressed the details of how positional information is established, neglecting potential issues with establishing a reliable gradient, and focused on the unavoidable impact of limited positional information in collective gradient sensing. 
Consequently, these results should in a sense be considered the best case for exploiting positional information: if the tradeoff argument of Fig. \ref{fig:figure4} predicts that maximum likelihood sensing using limited positional information is optimal, we should be skeptical about this. Contrarily, if the tradeoff suggests MLE is inferior to the tug-of-war, we should be highly confident that this is true. For this reason, it seems unlikely that collective gradient sensing by small clusters, such as border cell migration \cite{montell2012group,yue2018minimal,peercy2020clustered}, relies significantly on positional information.

Our results show that whether cells should use positional information in estimating a chemical gradient orientation depends not only on the amount of positional information but also on parameters like the cluster size and the gradient steepness.
Large clusters overcome positional uncertainties by having numerous cells, {\it i.e.} more independent measurements, combining their estimates of their positions and the local concentration of chemoattractant. 
This large benefit from gaining information from all cells and weighting them according to position can outweigh the costs of having limited positional information. This is especially apparent in comparison with the tug-of-war model, which only uses information from the small fraction of the cells at the cluster's edge.

Gradient steepness is also a key parameter. Finite positional information limits the accuracy of collective gradient sensing by a group of cells -- but this limitation becomes more significant for steeper gradients, since difference in positions lead to larger changes in the chemoattractant concentration.
Remarkably, for shallow gradients, ($g\rightarrow 0$), the limit where collective gradient sensing is most important, as single cells can become ineffective sensors, positional information turns out not to be relevant. 

While our simplest analytical results are derived for a cluster with a constant positional information $\Delta r$, we can generalize our results  beyond this point, finding that the effect of limited positional information differs depending on a cell's position within the cluster. 
Positional uncertainties in cells farther away from the cluster center of mass have a greater effect on the accuracy of gradient detection -- again, suggesting that edge cells must play a large role in sensing.
Our observations suggest that a group of cells may benefit from having specialized cells capable of sensing their position more accurately than the rest. The idea that cells may specialize depending on position in the cluster or chemoattractant concentration has also been suggested in other contexts \cite{hopkins2019leader,hopkins2020chemotaxis,copenhagen2018frustration,mclennan2015neural}.

In addition to the maximum likelihood estimation of \cite{camley2017cell}, which requires positional information, and the tug-of-war models \cite{camley2016emergent,malet2015collective,copenhagen2018frustration}, which do not, there are many other models of collective gradient sensing \cite{camley2018collective}. One broad category of a collective sensing mechanism that is not affected by positional information is collective local excitation--global inhibition (LEGI), where a local reporter reads out local chemoattractant concentration and a global inhibitor measures the global concentration~\cite{xiong2010cells, mugler2016limits,smith2016role,camley2016collective,varennes2016collective}. 
In LEGI, cell--cell communication imposes a maximum length scale at which gradient sensing information is reliably shared, leading to a saturation of gradient sensing accuracy with cell number~\cite{mugler2016limits}, compatible with experiments on mammary organoids \cite{ellison2016cell} but not lymphocyte clusters. 
While we have not treated the LEGI model in detail here, we argue that, similarly with the tug-of-war model, it does not require the cluster to know cell positions -- but it does not gain the benefits of measuring signals over the entire cluster. We would similarly expect LEGI sensing to be suboptimal to the maximum likelihood estimator for large clusters, shallow gradients, or low positional uncertainty.

Our results show that a sensing mechanism that does not rely on positional information, such as tug-of-war, is optimal for sufficiently small clusters or sufficiently large gradients. In Figure \ref{fig:figure4}, we have shown a phase diagram assuming that positional information is accurate to roughly a cell diameter, finding that tug-of-war is always optimal for clusters of seven cells, and will become optimal at large gradients for larger clusters. The experiments that originally suggested tug-of-war models where only edged cells participate in sensing and responding to the gradient~\cite{theveneau2010collective,malet2015collective} involved tens to hundreds of cells. If the tug-of-war behavior represents an evolutionary optimum, it may reflect either a low ability to gain positional information or a typical need to follow relatively steep gradients.

\section*{Acknowledgements}    
BAC acknowledges support from NSF grant PHY 1915491. We thank Kurmanbek Kaiyrbekov for a close reading of the manuscript. 

\appendix

\section{Derivation of the Fisher Information matrix and computing of the lower bound for the gradient sensing error through the maximum likelihood estimation method}
\label{sec:MLE_appendix}
We apply MLE to obtain the gradient sensing error of a group of cells that sense their local chemoattractant concentration and have limited positional information.
Each cell $i$ performs two independent measurements, a measurement of the local concentration $M_i$ and a measurement of its position $\vec{r}^*_i$ given by Eqs.~(\ref{eq:M_i}) and~(\ref{eq:ri}) respectively.
From the Results section, we know we can write the likelihood function as, $\mathcal{L}(\vec{g},\{\vec{r}_i\}|\{M_i, \vec{r}_i^*\})=\prod_i p(M_i|\vec{g},\vec{r}_i) p(\vec{r}_i^*|\vec{r}_i)$.
In general, it is easier to work with the log of the likelihood instead, since the product between the probabilities becomes a sum.
Then, considering a gradient $\vec{g} = g_x \hat{x}+g_y \hat{y}$ and a covariance matrix for the positional error, 
\begin{equation}
    \bm{\Sigma}_i = \left[
    \begin{matrix}
    \sigma_{x_i}^2 & \rho_i \sigma_{x_i}\sigma_{y_i} \\
    \rho_i \sigma_{x_i}\sigma_{y_i} & \sigma_{y_i}^2
    \end{matrix}
    \right],
\end{equation}
we find the log of the likelihood to be
\begin{widetext}
\begin{align}
    \label{eq:logL}
    \log \mathcal{L}(\vec{g}, \{\vec{r}_i\}|\{M_i, \vec{r}_i^*\}) =& -\frac{3}{2}\sum_i \log(2\pi)  -\frac{1}{2}\sum_i \log h_i - \sum_i \frac{(M_i - \mu_i)^2}{2h_i} -\frac{1}{2}\sum_i \log\left(\sigma_{x_i}^2\sigma_{y_i}^2(1-\rho_i^2)\right) \nonumber\\ 
     &-\sum_i\frac{1}{2(1-\rho_i^2)}\left[\frac{(r^*_{x_i}-r_{x_i})^2}{\sigma_{x_i}^2}-2\rho_i\frac{(r^*_{x_i}-r_{x_i})(r^*_{y_i}-r_{y_i})}{\sigma_{x_i}\sigma_{y_i}} +\frac{(r^*_{y_i}-r_{y_i})^2}{\sigma_{y_i}^2}\right].
\end{align}
\end{widetext}
where $\mu_i = 1 + \vec{g}\cdot\delta\vec{r}_i$ and $h_i = (\delta c_i/c_0)^2 + \sigma^2_\Delta$ are the mean and variance of the local concentration measurement performed by cell $i$.

From Eq.~(\ref{eq:logL}), we could find the global maximum to obtain the maximum likelihood estimator. 
Here, we are not interested in the estimator itself but in its fluctuations. 
We want to compute the estimator error, which we know must converge to the Cram\'er-Rao bound, given by the inverse of the Fisher information matrix, $\left(\mathcal{I}\right)^{-1}$. This bound limits the accuracy of {\it any} unbiased measurement of the gradient $\vec{g}$ \cite{kay1993fundamentals}.
This bound puts a limit on minimal errors of any unbiased estimator of a parameter (e.g. an estimator $\hat{\alpha}$) away from the true value of that parameter ($\alpha$). This bound is, for parameters $\alpha$ and $\beta$,
\begin{align}
    \langle (\alpha-\hat{\alpha})(\beta-\hat{\beta})\rangle = \left(\mathcal{I}^{-1}\right)_{\alpha,\beta},
    \label{eq:def_error_fisher_inf}
\end{align}
Recalling the Fisher information definition, $\mathcal{I}_{\alpha,\beta} = -\left\langle \frac{\partial^2\ln\mathcal{L}}{\partial \alpha\partial \beta}\right\rangle$, we next take partial derivatives of Eq.~(\ref{eq:logL}), and compute the expectation values, getting
\begin{widetext}
\begin{align*}
\resizebox{.99 \textwidth}{!}{$
\mathcal{I} =
\begin{bmatrix} 
\sum_i f_i\delta r_{x_i}^2  & \sum_i f_i\delta r_{x_i}\delta r_{y_i} & g_xf_1\delta r_{x_1}  &  g_yf_1\delta r_{x_1} & \dots & g_xf_N\delta r_{x_N}  &  g_yf_N\delta r_{x_N} \\
\sum_i f_i\delta r_{x_i}\delta r_{y_i} & \sum_i f_i\delta r_{y_i}^2 &
g_xf_1\delta r_{y_1} & g_yf_1\delta r_{y_1} & \dots & g_xf_N\delta r_{y_N} & g_yf_N\delta r_{y_N} \\
g_xf_1\delta r_{x_1} & g_xf_1\delta r_{y_1} &  g_x^2f_1 + \frac{m_1}{\sigma_{x_1}^2} & g_xg_yf_1 - \frac{m_1\rho_1}{\sigma_{x_1}\sigma_{y_1}}& \dots & 0 & 0 \\
g_yf_1\delta r_{x_1} & g_yf_1\delta r_{y_1} & g_xg_yf_1 - \frac{m_1\rho_1}{\sigma_{x_1}\sigma_{y_1}}&  g_y^2f_1 + \frac{m_1}{\sigma_{y_1}^2} & \dots & 0 & 0\\
\vdots & \vdots & \vdots & \vdots & \ddots & \vdots & \vdots\\
 g_xf_N\delta r_{x_N} & g_xf_N\delta r_{y_N} & 0 & 0 & \dots &  g_x^2f_N +\frac{m_N}{\sigma_{x_N}^2} & g_xg_yf_N - \frac{m_N\rho_N}{\sigma_{x_N}\sigma_{y_N}}\\
g_yf_N\delta r_{x_N} & g_yf_N\delta r_{y_N} & 0 & 0 & \dots & g_xg_yf_N - \frac{m_N\rho_N}{\sigma_{x_N}\sigma_{y_N}}& g_y^2f_N + \frac{m_N}{\sigma_{y_N}^2}
\end{bmatrix}
$}
\end{align*}
\end{widetext}
where $f_i = \frac{(a + \mu_i)^2\left((a+3\mu_i)^2 + 2 a n_r\mu_i\right) + 2a^2n_r^2\sigma_\Delta^2}{2\left(\mu_i(\mu_i+a)^2+ an_r\sigma_\Delta^2\right)^2}$, $m_i = \frac{1}{(1-\rho_i^2)}$, $a= K_D/c_0$, and $\delta \vec{r}_i = \vec{r}_i - \vec{r}_\mathrm{cm}$.
The order of the Fisher information matrix elements correspond to: $\{g_x, g_y, r_{x_1}, r_{y_1}, \dots, r_{x_N}, r_{y_N}\}$, such that, as for example, $\mathcal{I}_{1,1} = -\left\langle \frac{\partial^2\ln\mathcal{L}}{\partial g_x^2}\right\rangle$, $\mathcal{I}_{1,2} = -\left\langle \frac{\partial^2\ln\mathcal{L}}{\partial g_x\partial g_y}\right\rangle$, $\mathcal{I}_{1,2(N+1)} = -\left\langle \frac{\partial^2\ln\mathcal{L}}{\partial g_x\partial r_{y_N}}\right\rangle$.  
The derivation of the Fisher information matrix from the likelihood function is not hard, but takes some calculation effort.
To gain some intuition about the shape it takes, we point out some aspects. 
Note that $\vec{g}$ and $\vec{r}_i$ are only related by the likelihood function through the term $\vec{g}\cdot\delta\vec{r}_i$, which appears inside the expressions of $h_i$ and $\mu_i$.
This explains that the $f_i$ parameters are in every element of the Fisher information matrix, since taking two times the derivatives with respect to $\vec{g}$ or $\vec{r}_i$ leads to the same factor. 
Also, from the same term we can understand the full first two rows/columns in the Fisher information matrix, since these correspond to the derivatives with respect to $\vec{g}$ and $\vec{r}_i$.
Lastly, note that the elements related to the double derivatives with respect to positions have a contribution coming from the gradient terms, plus the contribution from the positional uncertainty terms of the likelihood function.

We are interested in computing the gradient sensing errors, $\sigma^2_{g_x} \equiv \langle\left( g_x - \hat{g}_x\right)^2\rangle = \left(\mathcal{I}^{-1}\right)_{g_x,g_x}$ and $\sigma^2_{g_y} \equiv \langle\left( g_y - \hat{g}_y\right)^2\rangle= \left(\mathcal{I}^{-1}\right)_{g_y,g_y}$. 
Following a guided Gaussian elimination procedure we can obtain the inverse of the Fisher information matrix for the elements associated with the gradient (see Appendix~\ref{sec:app_proof_mle}). 
The maximum likelihood estimator errors for the gradient are,
\begin{align}
    \sigma_{g_x}^2 =& \frac{1}{\mathcal{S}}\sum_{i} \gamma_i\delta r_{y_i}^2, \label{eqs:mle_general_solution_x}\\
    \sigma_{g_y}^2 =& \frac{1}{\mathcal{S}}\sum_{i} \gamma_i\delta r_{x_i}^2, \label{eqs:mle_general_solution_y}\\
    \sigma_{g_x,g_y} =&- \frac{1}{\mathcal{S}}\sum_{i} \gamma_i\delta r_{x_i}\delta r_{y_i},\label{eqs:mle_general_solution_xy}
\end{align}
where $\mathcal{S} = \sum_{i} \gamma_i\delta r_{x_i}^2\sum_{i} \gamma_i\delta r_{y_i}^2 - \Big(\sum_{i} \gamma_i\delta r_{x_i}\delta r_{y_i}\Big)^2$, and $\gamma_i = \frac{f_i}{1 + f_i\vec{g}^T\bm{\Sigma}_i\vec{g}}$.

Finally, it is easier to interpret the gradient sensing errors in terms of the steepness, $g$, and direction $\phi$, of the gradient. 
We can obtain them by re-parametrizing the Fisher information matrix as,
\begin{align*}
    I_{\vec{\theta'}} = \vec{J}^T I_{\vec{\theta}}(\vec{\theta}(\vec{\theta'}))\vec{J},
\end{align*}
where $\vec{J}$ is the Jacobian matrix, $J_{ij} =\frac{\partial \theta_i}{\partial \theta_j'}$, and $\theta = \{g_x, g_y\}$ and $\theta' = \{g, \phi\}$ are the variables we want to transform. 
The errors for the gradient steepness and direction then are,
\begin{align}
\label{eq:sigma_g_rotation_to_polar_coordinates}
    \sigma^2_g &= \cos^2(\phi)\sigma^2_{g_x} +  \sin^2(\phi)\sigma^2_{g_y} + 2\cos(\phi)\sin(\phi)\sigma_{g_x,g_y}\\
    \sigma^2_\phi &= \frac{\sin^2(\phi)\sigma^2_{g_x} +  \cos^2(\phi)\sigma^2_{g_y} - 2\cos(\phi)\sin(\phi)\sigma_{g_x,g_y}}{g^2}.\label{eq:sigma_phi2}
\end{align}
We presented these equations for an isotropic cluster in the main text, where $\sigma_{g_x,g_y} \approx 0$ and $\sigma_{g_x} \approx \sigma_{g_y}$. In this case, we find that $\sigma_\phi^2 = \sigma_g^2 / g^2$ -- but this is only true for isotropic clusters.
Note that Eq.~(\ref{eq:sigma_phi2}) fails for large enough $\sigma_g$, since $\phi$ is constrained between $0$--$2\pi$, and consequently, $\sigma_\phi$ is bounded. 

\subsection*{Shallow gradient approximation}

In the limit of shallow gradient approximation it is possible to obtain simpler expression for the gradient sensing errors. 
In this limit, $gR_{\mathrm{cluster}}\ll 1$, meaning that the cell's ligand-receptor fluctuations can be approximated as $\delta c_i \approx \overline{\delta c} = \sqrt{ \frac{1}{n_r} \frac{ c_0(c_0 + K_D)^2}{K_D}}$. %
Variations of the chemoattractant concentration along the cell's cluster are small, thus we can assume that all cells have the same reading concentration errors.
Then, $h_i \approx \bar{h} = \overline{\delta c}^2 + \sigma^2_\Delta$. 
Moreover, since $n_r\gg K_D/c_0$, $f_i$ can be approximated as the inverse of the reading concentration error, $f_i\approx \bar{f}=1/\bar{h}$.
Note that the limit of large cell-cell variability, $\sigma_\Delta\gg \delta c_i/c_0$, would lead to the same approximation.
Assuming constant isotropic positional errors, such that $\rho_i = 0$ and $\sigma_{x_i} = \sigma_{y_i} = \Delta r$, we find that $\gamma_i = \frac{1}{1/f_i + \vec{g}^T\bm{\Sigma}_i\vec{g}} \approx \frac{1}{\bar{h} + g^2 \Delta r^2}$ is independent of $i$ and thus  Eqs.~(\ref{eqs:mle_general_solution_x})--(\ref{eqs:mle_general_solution_xy}) take a simplified form,
\begin{align*}
    \sigma^2_{g_x} &= \frac{\sum_i \delta r_{y_i}^2}{\bar{\mathcal{S}}}\, \left(\bar{h} + g^2\Delta r^2\right),\\
    \sigma^2_{g_y} &= \frac{\sum_i \delta r_{x_i}^2}{\bar{\mathcal{S}}}\, \left(\bar{h} + g^2\Delta r^2\right),
\end{align*}
and in polar coordinates, 
\begin{align*}
    \sigma^2_g &= \Gamma_g \left(\bar{h} + g^2\Delta r^2\right),\\
    \sigma^2_\phi &= \frac{\Gamma_\phi}{g^2} \left(\bar{h} + g^2\Delta r^2\right),
\end{align*}
where $\Gamma_g = \frac{\sum_i (\sin(\phi)\delta r_{x_i} - \cos(\phi) \delta r_{y_i})^2}{\bar{\mathcal{S}}}$, $\Gamma_\phi = \frac{\sum_i (\cos(\phi)\delta r_{x_i} + \sin(\phi) \delta r_{y_i})^2}{\bar{\mathcal{S}}}$, and $\bar{\mathcal{S}} = \sum_i \delta r_{x_i}^2\sum_i \delta r_{y_i}^2 - (\sum_i \delta r_{x_i}\delta r_{y_i})^2$ are geometrical factors that depend on the cluster's shape and the gradient's orientation. 
Note that in the case the cluster have circular symmetry, ($\sum_i \delta r_{x_i}^2\approx \sum_i\delta r_{y_i}^2\approx \frac{1}{2}\sum_i\vert\delta\vec{r}_i\vert^2$, $\sum_i \delta r_{x_i}\delta r_{y_i}\approx 0$), then $\Gamma_g\approx \frac{1}{\chi}$, and Eq.~(\ref{eq:sigma_g2_SGA}) is recovered. 

\section{Auxiliary computations for the tug--of--war model} \label{sec:app_tug_of_war}
\subsection*{Details on computation of the mean and covariance of $\vec{p}_i(t)$} 
We want to compute the mean value and variance of $\vec{p}_i$. 
We start from Eq.~(\ref{eq:pi_sol}) .
The mean value is straight forward, 
\begin{widetext}
\begin{align*}
    \langle \vec{p}_i\rangle =& \Big\langle \bm{\mu}_i + \sigma_p\int_0^tdt' e^{-(t-t')/\tau_p}\bm{\epsilon}_i(t') + \bar{\beta}c_0\vec{q}_i\int_0^t dt' e^{-(t-t')/\tau_p}\Xi_i(t')\Big\rangle,\\
  = &\langle\bm{\mu}_i\rangle + \sigma_p\int_0^t dt' e^{-(t-t')/\tau_p}\langle\bm{\epsilon}_i(t')\rangle + \bar{\beta}c_0\vec{q}_i\int_0^t dt' e^{-(t-t')/\tau_p}\langle\Xi_i(t')\rangle,\\
 =& \bm{\mu}_i.
\end{align*}
\end{widetext}

Recalling that $\langle \Xi_i(t)\Xi_j(0)\rangle = \left(\delta c_i^2/c_0^2 C_c(t)+\sigma_\Delta^2 C_\Delta(t)\right)\delta_{ij}$, and that we have assumed that correlations decay exponentially, so $C_{s}(t) = e^{-t/\tau_s}$, with $\tau_s$ the correlation time and the index $s=\{c,\Delta\}$, we now compute the covariance of $\vec{p}_i(t)$,
\begin{widetext}
\begin{align*}
\left\langle (p_{\mu_i}(t)- \langle p_{\mu_i}\rangle)(p_{\nu_i}(t)-\langle p_{\nu_i}\rangle)\right\rangle =&
\,\, \sigma^2\int_0^t dt'\int_0^t dt'' \langle\epsilon_{\mu_i}(t')\epsilon_{\nu_i}(t'')\rangle e^{-(t-t')/\tau_p}e^{-(t-t'')/\tau_p} \\
&\,\, + \bar{\beta}^2c_0^2 q_{\mu_i}q_{\nu_i}\int_0^t dt'\int_0^t dt'' \langle\Xi_i(t')\Xi_i(t'')\rangle e^{-(t-t')/\tau_p}e^{-(t-t'')/\tau_p}\\
=&\,\, \frac{\sigma^2\tau_p}{2} + \bar{\beta}^2q_{\mu_i}q_{\nu_i}\delta c_i^2 \mathrm{T}(t,\tau_c) + \bar{\beta}^2c_0^2 q_{\mu_i}q_{\nu_i}\sigma_\Delta^2 \mathrm{T}(t,\tau_\Delta)\\
\end{align*}
\end{widetext}
where,
\begin{align*}
    \mathrm{T}(t,\tau_s) =& \frac{\tau_p^2}{1+\upsilon_s}\Big(1 + \frac{1}{1-\upsilon_s}\\
    &\times\big((1+\upsilon_s)e^{-2t/\tau_p} - 2e^{-(1+\upsilon_s)t/\tau_p}\big)\Big),
\end{align*}
and $\upsilon_s = \frac{\tau_p}{\tau_s}$. 
$\mathrm{T}(t,\tau_s)/\tau_p^2$ represent the time averaging function for a fluctuating process with correlation time $\tau_s$ up to a time $t$.
Given that $\tau_\Delta \gg t \gg \tau_p \gg \tau_c$, then $\mathrm{T}(t,\tau_c) \approx \frac{\tau_p^2}{1+\tau_p/\tau_c}\approx \tau_p\tau_c$ and $\mathrm{T}(t,\tau_\Delta) \approx \tau_p^2$, so then $\mathrm{T}(t,\tau_\Delta) \gg \mathrm{T}(t,\tau_c)$.

\subsection*{Error propagation for the tug-of-war model}

To estimate the gradient sensing error for the tug--of--war model we use a simple error propagation method. 
From Eq.~(\ref{eq:v_cluster_mean}), we can write an expression for the gradient related to the cluster velocity, $\vec{g}=\frac{1}{\gamma c_0}\mathcal{M}^{-1}\langle \vec{v}_c\rangle$, where,
\begin{equation}
    \mathcal{M}^{-1} = \frac{N}{\mathcal{S}_\vec{v}} \begin{bmatrix}
     \sum_i\delta r_{y_i}q_{y_i} &  -\sum_i\delta r_{y_i}q_{x_i}\\
     -\sum_i\delta r_{x_i}q_{y_i} &  \sum_i\delta r_{x_i}q_{x_i}\\
    \end{bmatrix},
\end{equation}
and $\mathcal{S}_\vec{v} = \sum_i\delta r_{x_i}q_{x_i}\sum_i\delta r_{y_i}q_{y_i} - \sum_i\delta r_{x_i}q_{y_i}\sum_i\delta r_{y_i}q_{x_i}$.

Recalling the symmetry of the roughly circular hexagonal clusters, the following relations are fulfilled, $\sum_i \delta r_{x_i}^2\approx \sum_i\delta r_{y_i}^2 \approx \frac{1}{2}\sum_i\vert\delta\vec{r}_i\vert^2$ , $\sum_i \delta r_{x_i}\delta r_{y_i}\approx 0$, $\sum_i q_{x_i}^2\approx \sum_i q_{y_i}^2\approx \frac{1}{2}\sum_i\vert\vec{q}_i\vert^2$, $\sum_i q_{x_i}q_{y_i}\approx 0$, $\sum_i q_{x_i}\delta r_{x_i} \approx \sum_i q_{y_i}\delta r_{y_i}\approx \frac{1}{2}\sum_i \vec{q}_i\cdot\delta \vec{r}_i$, and $\sum_i q_{x_i}\delta r_{y_i}\approx \sum_i q_{y_i}\delta r_{x_i}\approx 0$. 
In addition, note that in such a case, we can write $\mathcal{S}_\vec{v} \approx \sum_i\delta r_{x_i}q_{x_i}\sum_i\delta r_{y_i}q_{y_i} \approx \frac{1}{4}\left(\sum_i \vec{q}_i\cdot\delta \vec{r}_i\right)^2$ and,
\begin{equation}
\mathcal{M}^{-1} \approx \frac{2N}{\sum_i \vec{q}_i\cdot\delta \vec{r}_i} \mathbb{I},
\end{equation}
where $\mathbb{I}$ is the $2\times2$ identity matrix.
Then, the expression for the gradient takes the following simple form, 
\begin{equation}
\label{eq:g_vs_vc_symmetry}
    \vec{g} = \frac{2N}{\gamma c_0 \sum_i \vec{q}_i\cdot\delta \vec{r}_i} \langle\vec{v}_c\rangle.
\end{equation}
Eq.~(\ref{eq:g_vs_vc_symmetry}) shows that $\vec{g}$ and $\langle \vec{v}_c\rangle$ are proportional to each other.
Then, we can write the variance of $\vec{g}$ as,
\begin{equation}
\label{eq:var_g_vs_var_vc}
    \mathrm{var}(\vec{g}) = \left(\frac{2N}{\gamma c_0 \sum_i \vec{q}_i\cdot\delta \vec{r}_i}\right)^2 \mathrm{var}(\vec{v}_c).
\end{equation}
Introducing the velocity variance expressions given by Eq.~(\ref{eq:tug-of-war_v_cluster_errors}) in the main text into Eq.~(\ref{eq:var_g_vs_var_vc}), we finally obtain the gradient sensing variances for the tug--of--war model, 
\begin{align}
\label{eq:gradient_sensing_errors_xy_tug_of_war}
    \sigma_{g_x}^2 \approx \sigma^2_{g_y} \approx &\frac{2}{\left(\sum_i \vec{q}_i\cdot\delta \vec{r}_i\right)^2}  \left(\sum_i \vert \vec{q}_i\vert^2 h_{Ti}+ \frac{\sigma_p^2\tau_p}{\gamma^2c_0^2}N \right),\\
    \mathrm{cov}(g_x,g_y) \approx& 0.
\end{align}

We note that, though our derivation of this equation does involve a sum over cell positions, the cluster can compute the direction of the gradient through the tug-of-war {\it without} explicitly knowing any cell locations.

\section{Maximum likelihood estimator with time averaging}
\label{sec:mle_time_average}
In the main text we apply the MLE method to compute the gradient sensing error from the individual cell measurements of the chemoattractant concentration, $M_i$, and cell positions, $\vec{r}^*_i$. These measurements capture a ``snapshot'' or instant of the cluster state. 
Later in the text, we compare the MLE outcomes with the tug--of--war model, which relies on the statistics of the cell polarities $\vec{p}_i$ obtained for asymptotic times larger than $\tau_p$. 
Therefore, we are comparing instantaneous with time averaged measurements. 
We could do this since the sensing concentration measurements are dominated by the CCV which has a correlation times larger than the polarity relaxation time. 
Here we extend the MLE method by using time average measurements in the same way as the tug--of--war model.

We compute the time average for the individual cell chemoattractant concentration measurements, 
\begin{align*}
    M_i^T(t) &= \int M_i(t')K_T(t-t')dt',\\
     &= c(\vec{r}_i)/c_0 + \int \Xi_i(t')K_T(t-t')dt',
\end{align*}
where $K_T(t) = \mathcal{H}(t)\frac{1}{T}e^{-t/T}$ is a time averaging kernel and $\mathcal{H}(t)$ the Heaviside function.
We assume that cell positions remain fixed during the averaging time $T$, so that $c(\vec{r}_i,t)=c(\vec{r}_i)$.
Now we compute the mean and variance for $M_i^T$.
\begin{align*}
    \langle M_i^T\rangle = c(\vec{r}_i)/c_0,
\end{align*}

\begin{widetext}
\begin{align*}
    \left\langle \left(M_i^T - \langle M_i^T\rangle\right)^2\right\rangle &= \frac{1}{T^2} \int_{0}^t dt'\int_{0}^t dt'' \langle\Xi_i(t')\Xi_i(t'')\rangle e^{-(t-t')/T}e^{-(t-t'')/T},\\
    &= \frac{1}{T^2} \int_{0}^t dt'\int_{0}^t dt''\left(\frac{\delta c_i^2}{c_0^2}C_{c}(\vert t'-t''\vert) + \sigma_\Delta^2C_{\Delta}(\vert t'-t''\vert)\right) e^{-(t-t')/T}e^{-(t-t'')/T},\\
    &= \frac{1}{1+T/\tau_c}\frac{\delta c_i^2}{c_0^2} + \frac{1}{1+T/\tau_\Delta}\sigma_\Delta^2.
\end{align*}
\end{widetext}

If we substitute the averaging time $T = \tau_p$, these are precisely the values $h_{T_i}$ which we derived in our tug-of-war model. This shows we can carry over all of our results from the maximum likelihood estimation to match the results of the tug-of-war model. However, this is not particularly important, because $T\gg\tau_c$ and $T\ll\tau_\Delta$, the sensing concentration fluctuations can be approximated by $\langle \left(M_i^T - \langle M_i^T\rangle\right)^2\rangle\approx \sigma_\Delta^2$. 
When we take into account time averaging, the contributions to the sensing concentration fluctuations are dominated by CCV while ligand-receptor fluctuations are averaged out.

\section{Proof for the analytical solution of the MLE errors} \label{sec:app_proof_mle}
In this appendix we show the derivation of the expressions for the MLE errors in the shallow gradient limit. 
The problem we are solving is very similar to an error--in--variable models of regression, in which the regressor variable is subject to errors~\cite{fuller2009measurement}.
However, we were not able to find an explicit mapping between these results and our question of interest -- computing the Fisher information matrix and its inverse specifically for the $g_x$ and $g_y$ variables.
Finding the Fisher information matrix $\mathcal{I}$ is relatively straightforward from evaluating derivatives of the log-likelihood, and it is presented in Appendix \ref{sec:MLE_appendix}.
However, to determine the Cramer-Rao bound, we need to compute the {\it inverse} of the Fisher information matrix -- in principle needing to analytically invert a matrix whose size scales with the number of cells. This is nontrivial in general. However, for this problem we were able to 
use Gaussian elimination to find the elements of the inverse of the Fisher information matrix necessary for us, {\it i.e.} $\left(\mathcal{I}\right)^{-1}_{g_xg_x}$, $\left(\mathcal{I}\right)^{-1}_{g_yg_y}$, and $\left(\mathcal{I}\right)^{-1}_{g_xg_y}$.
Given the almost tridiagonal shape of the Fisher matrix, we first show the procedure for how to eliminate the elements of the rows associated with a individual cell $s$ and then repeat the same steps for the rest of the cells.

Recall that the Gaussian elimination method is a linear algebra method to find the solution of a problem $\mathcal{A}\times X=\mathcal{B}$. 
This problem can be represented in terms of an augmented matrix $[\mathcal{A}\vert\mathcal{B}]$. Row operations (i.e. linear combinations of the rows) are performed, leading eventually to finding the solution for $X$ when the augmented matrix takes the form $\left[\mathbb{I}\vert\mathcal{C}\right]$, where $\mathbb{I}$ is the identity matrix, and $\mathcal{C}= \left(\mathcal{A}^{-1}\right)\mathcal{B}=X$. 
In our case, the problem we want to solve is to find the inverse of the Fisher information matrix, $\mathcal{I}\times\left(\mathcal{I}^{-1}\right) = \mathbb{I}$.
To start with the Gaussian elimination process, we first write the augmented matrix $\left[\mathcal{I}\vert\mathbb{I}\right]$,
\begin{widetext}
\begin{align*}
\left[\begin{matrix} 
S_{xx}  & S_{xy} & \dots & g_xf_s\delta r_{x_s}  &  g_yf_s\delta r_{x_s} & \dots \\
S_{xy} & S_{yy} & \dots &
g_xf_s\delta r_{y_s} & g_yf_s\delta r_{y_s} & \dots \\
\vdots & \vdots & \dots & \vdots & \vdots & \dots \\
g_xf_s\delta r_{x_s} & g_xf_s\delta r_{y_s} & \dots & g_x^2f_s + \frac{m_s}{\sigma_{x_s}^2} & g_xg_yf_s + \frac{m_s\rho_s}{\sigma_{x_s}\sigma_{y_s}} & \dots \\
g_yf_s\delta r_{x_s} & g_y f_s\delta r_{y_s} & \dots & g_xg_yf_s + \frac{m_s\rho_s}{\sigma_{x_s}\sigma_{y_s}}&  g_y^2f_s +  \frac{m_s}{\sigma_{y_s}^2} & \dots \\
\vdots & \vdots & \dots & \vdots & \vdots & \dots\\
\end{matrix}\right\vert
\left.\begin{matrix}
1 & 0 & \dots\\
0 & 1 & \dots\\
\vdots & \vdots & \dots\\
0 & 0 & \dots\\
0 & 0 & \dots\\
\vdots & \vdots & \dots\\
\end{matrix}\right]
\end{align*}
\end{widetext}
where $S_{xx} = \sum_if_i\delta r_{x_i}^2$, $S_{yy} = \sum_if_i\delta r_{y_i}^2$, and $S_{xy}= \sum_if_i\delta r_{x_i}\delta r_{y_i}$.
Since we are only looking for the gradient estimation errors, we solely consider the two first columns of the right matrix.

Following, we apply the operations to perform an upper diagonalization of columns associated to the cell $s$. 
From now on we use the notation $R^{i}$ and $\mathcal{I}_{m,n}$ to identify the $i$--row and the $(m,n)$--element of the matrix respectively. 
Note that the rows associated with the cell $s$ are $R^{2s+1}$ and $R^{2s+2}$. 

\begin{enumerate}
    \item Set element $\mathcal{I}_{2s+2,2s+2}=1$: $R^{2s+2} \leftarrow \frac{R^{2s+2}}{\mathcal{I}_{2s+2,2s+2}}$.
    \item Set element $\mathcal{I}_{2s+1,2s+2}=0$: $R^{2s+1} \leftarrow R^{2s+1} - \mathcal{I}_{2s+1,2s+2}R^{2s+2}$.
    \item Set element $\mathcal{I}_{2,2s+2}=0$: $R^{2} \leftarrow R^{2} - \mathcal{I}_{2,2s+2}R^{2s+2}$.
    \item Set element $\mathcal{I}_{1,2s+2}=0$: $R^1 \leftarrow R^1 - \mathcal{I}_{1,2s+2}R^{2s+2}$.
    \item Set element $\mathcal{I}_{2s+1,2s+1}=1$: $R^{2s+1} \leftarrow \frac{R^{2s+1}}{\mathcal{I}_{2s+1,2s+1}}$.
    \item Set element $\mathcal{I}_{2,2s+1}=0$: $R^2 \leftarrow R^2 - \mathcal{I}_{2,2s+1}R^{2s+1}$.
    \item Set element $\mathcal{I}_{1,2s+1}=0$: $R^1 \leftarrow R^1 - \mathcal{I}_{1,2s+1}R^{2s+1}$.
\end{enumerate}

After these steps we arrive to the following matrix,

\begin{align*}
\resizebox{.99\hsize}{!}{$
\left[\begin{matrix} 
S_{xx}-\alpha_s\delta r_{x_s}^2 & S_{xy}-\alpha_s\delta r_{x_s}\delta r_{y_s}& \dots & 0 &  0  & \dots \\
S_{xy} - \alpha_s\delta r_{x_s}\delta r_{y_s} & S_{yy}- \alpha_s\delta r_{y_s}^2 & \dots &
0 & 0 & \dots \\
\vdots & \vdots & \dots & \vdots & \vdots & \dots \\
\Box & \Box & \dots & 1 & 0 & \dots \\
\Box & \Box & \dots & \Box & 1 & \dots \\
\vdots & \vdots & \dots & \vdots & \vdots & \dots\\
\end{matrix}\right\vert
\left.\begin{matrix}
1 & 0 & \dots\\
0 & 1 & \dots\\
\vdots & \vdots & \dots\\
0 & 0 & \dots\\
0 & 0 & \dots\\
\vdots & \vdots & \dots\\
\end{matrix}\right]
$}
\end{align*}
$\alpha_s = f_s^2\frac{\vec{g}\bm{\Sigma_s}\vec{g}^T}{1 + f_s\vec{g}\bm{\Sigma_s}\vec{g}^T}$ and the squares ($\Box$) represent elements with large expressions that are not relevant for the computation. 

Then, we need to repeat steps 1--7 for all $s$. For each iteration we add a term $-\alpha_s \delta r_{z_ks}\delta r_{z_ls}$ to the matrix elements $\{k,l\}$ with $k,l = \{1,2\}$ and $z_{k,l} = \{x,y\}$. Finally, after repeating the procedure for all $s$, we get that the elements $\{k,l\}$ results in $\sum_i f_i \delta r_{z_xi}\delta r_{z_li} - \sum_s \alpha_s\delta r_{z_xs}\delta r_{z_ls} = \sum_i \gamma_i \delta r_{z_xi}\delta r_{z_li}$, 
where $\gamma_i = \frac{f_i}{1 + f_i\vec{g}\bm{\Sigma_i}\vec{g}^T}$, and the matrix reads,
\begin{align*}
\left[\begin{matrix} 
 \sum_i \gamma_i \delta r_{x_i}^2 & \sum_i \gamma_i \delta r_{x_i}\delta r_{y_i}  & \dots \\
\sum_i \gamma_i \delta r_{x_i}\delta r_{y_i} & \sum_i \gamma_i \delta r_{y_i}^2 & \dots \\
\vdots & \vdots & \dots\\
\end{matrix}\right\vert
\left.\begin{matrix}
1 & 0 & \dots\\
0 & 1 & \dots\\
\vdots & \vdots & \dots\\
\end{matrix}\right]
\end{align*}

Finally, since the first two rows are composed by a $2\times2$ block followed by all zeros to the right, instead of performing row operations, we can just invert this $2\times2$ matrix block to arrive to the final solution for the inverse of the Fisher matrix and the MLE errors. 
\begin{align*}
\left[\begin{matrix} 
1  & 0 & \dots \\
0 & 1 & \dots \\
\vdots & \vdots & \dots\\
\end{matrix}\right\vert
\left.\begin{matrix}
\frac{\sum_i \gamma_i \delta r_{y_i}^2}{\mathcal{S}} & -\frac{\sum_i \gamma_i \delta r_{x_i}\delta r_{y_i}}{\mathcal{S}} & \dots\\
-\frac{\sum_i \gamma_i \delta r_{x_i}\delta r_{y_i}}{\mathcal{S}} & \frac{\sum_i \gamma_i \delta r_{x_i}^2}{\mathcal{S}} & \dots\\
\vdots & \vdots & \dots\\
\end{matrix}\right]
\end{align*}
We can be positive that this is the final solution, since following with the Gaussian elimination process, no further operations over the first two rows would be necessary, and therefore, they will not be modified. 

We have ensured that this result is correct both by checking with computer algebra packages for small numbers of cells, as well as explicitly numerically inverting the Fisher information matrix.

\section{Default parameters}
In Table~\ref{tab:parameters} we show the default values for parameters in this study  with a brief justification of our choices. Values differing from these are indicated in the text and the figure captions.

\begin{table}[!h]
\begin{tabular}{ |p{0.2\linewidth}|p{0.25\linewidth}|p{0.45\linewidth}| } 
 \hline
 parameters & value & justification \\ 
 \hline
 $Q$ & $3$ & $N = 8-100$. From \cite{malet2015collective} area of clusters is $\sim 2500 \mu\mathrm{m}^2$ and considering that $R_\mathrm{cell} \sim 3-10 \mu\mathrm{m}$.\\ 
 $R_\mathrm{cell}$ & $10\mu$m & Lymphocytes $\sim 3-5\mu$m, epithelial cells $\sim 4-10 \mu$m.\\
 $n_r$ & $10^{5}$ & From \cite{hesselgesser1998identification, macdonald2008heterogeneity}.\\
 $a = \frac{K_d}{c_0}$ & $1$ & Optimal setting for reducing ligand-receptor noise~\cite{camley2017cell}.\\
 $\Delta r$ & $2R_\mathrm{cell}$ (1 cell) & Error estimation in \textit{Drosophila m.}  embryos ~\cite{gregor2007probing}\\
 $g_0$ & $0.005 \mu\mathrm{m}^{-1}$ & According to~\cite{camley2017cell} and \cite{malet2015collective}.\\
 $\sigma_{\Delta}$ & $0.1$ & According to~\cite{niepel2009non}\\
 $\sigma_p$ & $0$ & No polarization error. Best estimation the tug-of-war model can make. \\
 \hline
\end{tabular}
\caption{Parameter selection. Unless otherwise indicated in the text, these are the parameters used throughout the results. }
\label{tab:parameters}
\end{table}

\section*{References}
\bibliographystyle{unsrt}
\bibliography{biblio}

\end{document}